\documentclass[%
 reprint,
 superscriptaddress,
 amsmath,amssymb,
 aps,
prb,
]{revtex4-1}

\usepackage{graphicx}
\usepackage{subfigure}
\renewcommand{\thesubfigure}{\alph{subfigure}}
\makeatletter
\renewcommand{\@thesubfigure}{(\thesubfigure)\space}
\makeatother
\usepackage{dcolumn}
\usepackage{bm}
\usepackage[colorlinks, linkcolor=blue, citecolor=blue, urlcolor=blue]{hyperref}

\begin{document}

\preprint{APS/123-QED}

\title{Influence of hydrogen on electron-phonon coupling and intrinsic electrical resistivity in zirconium: a first-principles study}

\author{Qicheng Tang}
 \email{tangqicheng@westlake.edu.cn}
\affiliation{%
Zhejiang University, No.38, Zheda Road, Hangzhou 310027, China
}
\affiliation{%
Westlake University, No.18, Shilongshan Road, Hangzhou 310024, China
}
\author{L. A. Svyatkin}

\author{I. P. Chernov}
\affiliation{%
Tomsk Polytechnic University, No.30, Lenina Avenue, Tomsk 634050, Russia
}%

\date{\today}

\begin{abstract}
This paper presents the first-principles calculation of the electron-phonon coupling and the temperature dependence of the intrinsic electrical resistivity of the zirconium-hydrogen system with various hydrogen concentrations. The nature of the anomalous decrease in the electrical resistivity of the Zr-H system with the increase of hydrogen concentration (at high concentrations of H/Zr$>$1.5) is studied. It was found that the hydrogen concentration, where the resistivity starts to decrease, is very close to the critical concentration of the $\delta$-$\varepsilon$ phase transition. It is shown that the tetragonal lattice distortion due to the $\delta$-$\varepsilon$ phase transition of the Zr-H system eliminates imaginary phonon frequencies and the strong electron-phonon coupling of the $\delta$ phase and, as a result, leads to the reduction of the electrical resistivity of the Zr-H system at a high hydrogen concentration.
\begin{description}

\item[PACS numbers]
63.20.Kr, 64.70.Kb, 71.20.Be, 72.15.-v

\end{description}
\end{abstract}

\pacs{Valid PACS appear here}
\maketitle


\section{\label{sec:level1}Introduction}

The features of the metal-hydrogen interaction have been extensively studied over many years \cite{mueller2013metal, sakintuna2007metal, fukai2006metal}. The addition of hydrogen in metals leads to the change in its mechanical, transport, magnetic and other properties caused by the electron density redistribution and the change in the phonon structure and electron-phonon coupling. 
%
One of the most interesting changes is that the hydrogen absorption leads to a significant increase of the superconducting critical temperature $T_c$, which was first found in thorium \cite{satterthwaite1970superconductivity} and in palladium \cite{skoskiewicz1972superconductivity}. The nature of the increasing $T_c$ is the high Debye temperature due to the light mass of hydrogen. Recently, the high-$T_c$ superconductivity in H$_3$S \cite{errea2015high, drozdov2015conventional} and LaH$_{10}$ \cite{peng2017hydrogen, drozdov2018superconductivity} was reported in both experimental and theoretical studies, which provided credible evidence of the possibility to build a room-temperature superconductor from BCS theory.

The electron-phonon coupling (EPC), known as an importan physical process in metal superconductivity, is also important for studying its other transport properties, in particular, electrical resistance.
The measurement of the electrical resistivity can help us to identify the type of defects (in particular impurities) and their concentration in a real material. Moreover, the data for electrical resistivity are also helpful for understanding the heat conduction and other thermal properties of metals. 
An interesting fact observed in the experiments is the reduction of the electrical resistivity in some hydrogen-metal systems at very high H concentrations. In the Pd-H system this reduction occurs at the concentration H/Pd$\approx$0.71 \cite{geerken1983concentration} at 300 K and H(D)/Pd$\approx$0.75 \cite{sakamoto1996electrical} at 298 K; in the Zr-H system it occurs at the concentration H/Zr$\approx$1.6 \cite{bickel1970electrical} at 300 K. The correlation between the reduction of the electrical resisitivity and the phase transition of the Pd-H and Zr-H systems was assumed in experiments \cite{geerken1983concentration, bickel1970electrical}, using empirical models; however, it has still not been clarified by theoretical calculation. The understanding of the experimental observation is still open.

This work focuses on the study of the Zr-H system. Zr-based alloys are extensively used as structural materials for active zones of light-water reactors since Zr has a low thermal-neutron absorption cross section and good corrosion resistance and strength characteristics. 
 Hydrogen is actively accumulated in the materials during the reactor's operation and causes their corrosion and degradation of their mechanical properties \cite{wang2012first, zielinski2011hydrogen, zuzek1990h}. 
To study the influence of hydrogen on the mechanical properties of zirconium-based alloys, knowledge of the atomic structure of the Zr-H system is necessary.
 It was shown in experimental and theoretical works that at low H concentrations, the Zr-H system has a hexagonal close packed (HCP) structure and H atoms are located at tetragonal interstitial sites \cite{khoda1982determination, narang1977location, lopatina2015electronic, svyatkin2018mutual}. 
 At high H concentrations  ($1\leqslant$ H/Zr $\leqslant 2$), the Zr-H system has a face-centered cubic (FCC) or face-centered tetragonal (FCT) structure \cite{zuzek1990h, ackland1998embrittlement, niedzwiedz199391zr, cantrell1984x, bowman1983electronic}. The transition from the $\delta$ phase (FCC structure) to the $\varepsilon$ phase (FCT structure, $c<a$) was investigated in Refs. \cite{yakel1958thermocrystallography, bowman1983electronic, niedzwiedz199391zr, kul1999electronic, miwa2002first, quijano2009electronic}, and it is known that the critical H concentration of this phase transition is located at $x\geqslant 1.5$. It has been reported that there is a strong reduction in the electron state density at Fermi level due to the $\delta-\varepsilon$ phase transition (for ZrH$_2$ this reduction achieves 0.5 state/eV per unit cell \cite{quijano2009electronic}). The electron-driven mechanisms of the $\delta-\varepsilon$ phase transition in the Zr-H system with $1\leqslant x \leqslant 2$ were investigated in the works \cite{gupta1982electronic, cantrell1984x, wolf2000first, quijano2009electronic, zhang2011first, wang2012mechanical}. 
It should be mentioned that the expected correlation between the phase transition and decreasing resistance implies the importance of the electron-phonon coupling to the $\delta-\varepsilon$ phase transition in the Zr-H system.  But, to our knowledge, the key factors of electron-phonon coupling, varying with the change in the hydrogen concentration, are still unclear.

The electrical properties of the Zr-H system were measured experimentally for different purposes, such as researching the hydrogen kinetics \cite{grib2010kinetics}, the thermal properties \cite{tsuchiya2002thermal}, and the isotope effect \cite{uno2004thermophysical, tsuchiya2002isotope}.
Unlike the Pd-H system, the phonon distribution in the Zr-H system can be described by linear-response theory since the hydrogen atoms in Zr behave like "Einstein" oscillators (independent simple harmonic oscillators). 
This makes the first-principles calculation of EPC in the Zr-H system possible within density functional perturbation theory (DFPT). 
In present work, the H concentration's influence on the electron-phonon coupling and intrinsic electrical resistivity of Zr is theoretically studied. The electron-phonon coupling is analyzed by means of the Eliashberg function $\alpha^2F\left(\omega\right)$ and the Eliashberg transport function $\alpha_{tr}^2F\left(\omega\right)$. To clarify the relationship between the $\delta-\varepsilon$ phase transition and the reduction of the electrical resistivity, the influence of the tetragonal lattice distortion on the electron-phonon coupling of ZrH$_2$, as an example, was investigated.
We study the main factors decreasing the resistance of the Zr-H system due to its $\delta-\varepsilon$ phase transition.
Before the calculation of the electron-phonon coupling, we have determinated the lattice structure of the Zr-H system, and the results are discussed in the Appendix~\ref{APP}.

\section{\label{sec:Method and computational details}Methodology and computational details}

All calculations in our work were carried out from first principles within density functional theory and DFPT using the optimized norm-conserving Vanderbilt pseudopotential method (ONCVPSP) \cite{hamann2013optimized}, as implemented in the ABINIT \cite{gonze2016recent} code. To describe the exchange and correlation effects, the generalized gradient approximation (GGA) in the form of Perdew, Burke, and Ernzerhof \cite{perdew1996generalized} was used.
The cut off energy for the plane wave basis was set to 30 Ha in the structural optimization and relaxation and 40 Ha in the electronic structure calculation and linear-response calculation. To describe the occupation of electron levels, the temperature smearing method of ``cold smearing" \cite{verstraete2002} with a broadening of 0.001 Ha (about 316 K) was adopted, and the cut off energy of the smearing function was set to 0.5 Ha. For structural optimization, the Broyden-Fletcher-Goldfarb-Shanno minimization \cite{eitan2011optimal} was adopted. The atoms in the system considered were assumed to be in the equilibrium configuration when the force on each atom was below $10^{-4}$ Ha/bohr.

The present calculations were performed for ZrH$_{0.5}$, ZrH, ZrH$_{1.25}$, ZrH$_{1.5}$, ZrH$_{1.75}$, ZrH$_2$ and pure Zr. In our calculations we considered three possible structures: HCP, FCC, and FCT, with the H atoms all in tetragonal (T) sites or all in octaherdral (O) sites (see the Appendix~\ref{APP}).
To carry out the structural optimization and relaxation of the system considered, a supercell with 4 Zr atoms was adopted, and the $k$ meshes were chosen to be $13\times13\times4$ for the HCP structure and $14\times14\times14$ for the FCC, the FCT structures. 
In the electronic structure calculations [including the calculation of a band structure and the electronic density of states (EDOS)] for the ZrH$_2$ $\delta$ and $\varepsilon$ phases, the $k$ meshes of $32\times32\times32$ and $30\times30\times34$, respectively, were adopted.
In linear-response calculations, for pure Zr and ZrH a supercell with two Zr atoms was adopted, and a $k$ mesh of $14\times14\times10$ and a $q$ mesh of $7\times7\times5$ were chosen. For ZrH$_{1.25}$, ZrH$_{1.5}$, and ZrH$_{1.75}$  the supercell with four Zr atoms was adopted, a $k$ mesh of $14\times14\times14$ and the q-mesh of $7\times7\times7$ were chosen. For ZrH$_2$ the primitive cell was used, a $k$ mesh of $24\times24\times24$ and the q-mesh of $12\times12\times12$ were chosen.

By means of DFPT within the linear-response theory, the first-order perturbation potentials $\Delta_{q \upsilon}v$ for a phonon with frequency $\omega_{\bm{q}\upsilon}$ (crystal momentum $\bm{q}$ and branch index $\upsilon$) were calculated. Then they were used in the calculation of the electron-phonon matrix elements as $g_{ij\nu}\left(\bm{k},\bm{q}\right)=\left\langle\psi_{i\bm{k}+\bm{q}}\middle|\Delta_{q \upsilon}v|\psi_{j\bm{k}}\right\rangle$.  More details about the theoretical methods can be found in the review by Giustino \cite{giustino2017electron}.
The Eliashberg function \cite{allen1972neutron}, which measures the contribution of the phonons with frequency $\omega$ to scattering processes of the electrons at the Fermi level, was calculated in terms of the phonon linewidths $\gamma_{\bm{q}\upsilon}$,
\begin{eqnarray} 
\alpha^2F\left(\omega\right)=\frac{1}{2\pi N(\varepsilon_F)} \sum_{\bm{q}\upsilon}{ \frac{\gamma_{\bm{q}\upsilon}}{\omega_{\bm{q}\upsilon}} \delta(\hbar\omega-\hbar\omega_{\bm{q}\upsilon})}\;,
\end{eqnarray} 
where $N(\varepsilon_F)$ is the EDOS per atom and spin at the Fermi level $\varepsilon_F$. The linewidth $\gamma_{\bm{q}\upsilon}$ is written by electron-phonon matrix elements $g_{ij\nu}\left(\bm{k},\bm{q}\right)$:
\begin{eqnarray}
\gamma_{\bm{q}\upsilon}=2\pi \omega_{\bm{q}\upsilon} {\sum_{ij\bm{k}}\left|g_{ij\nu}\left(\bm{k},\bm{q}\right)\right|^2\delta(\varepsilon_{j\bm{k}}-\varepsilon_F)\delta(\varepsilon_{i\bm{k}+\bm{q}}-\varepsilon_F)}\;.& \nonumber \\ 
\end{eqnarray} 
The strength of $\alpha^2F\left(\omega\right)$ is described by the parameter
\begin{eqnarray}
\lambda=2\int_{0}^{\infty}{\frac{d\omega}{\omega}\alpha^2F\left(\omega\right)}\;,
\end{eqnarray} 
which is called the electron-phonon coupling constant.

For the calculation of transport properties, the tetrahedron smearing method \cite{zaharioudakis2004tetrahedron} was adopted. The Eliashberg transport function \cite{allen1971electron}, which is used to describe the influence of the electron-phonon scattering on transport properties, and additionally considers the efficiency factor of electronic transport $\eta_{\substack{\bm{k},\bm{q}\\{ij\nu}}}$, is calculated as follows
\begin{eqnarray}
\alpha_{tr}^2F\left(\omega\right)&=&\frac{1}{N(\varepsilon_F)}\sum_{\bm{q}\upsilon}{\sum_{ij\bm{k}}\eta_{\substack{\bm{k},\bm{q}\\{ij\nu}}}\left|g_{ij\nu}\left(\bm{k},\bm{q}\right)\right|^2\delta(\varepsilon_{j\bm{k}}-\varepsilon_F)} \nonumber \\ & &\times{\delta(\varepsilon_{i\bm{k}+\bm{q}}-\varepsilon_F)\delta(\hbar\omega-\hbar\omega_{\bm{q}\upsilon})}\;,
\end{eqnarray}
where an efficiency factor 
$\eta_{\substack{\bm{k},\bm{q}\\{ij\nu}}}=1-\frac{\bm{v}_{i\bm{k}+\bm{q}}\cdot \bm{v}_{j\bm{k}}}{\left \langle \bm{v}^2 \right \rangle}$
has been introduced in the terms of the electron velocity $\bm{v}_{j\bm{k}}$ in the state $\left.\left|\psi_{j\bm{k}}\right.\right\rangle$, with $\left \langle \bm{v}^2 \right \rangle$ being the average square of the Fermi velocity. The strength of $\alpha_{tr}^2F\left(\omega\right)$ is described by the parameter
\begin{eqnarray}
\lambda_{tr}=2\int_{0}^{\infty}{\frac{d\omega}{\omega}\alpha_{tr}^2F\left(\omega\right)}\;,
\end{eqnarray}
which is called the transport constant. For a metal, the electrical resistivity can be calculated by solving the Boltzmann equation in the lowest-order variational approximation \cite{savrasov1996electron}, and can be written in terms of $\alpha_{tr}^2F\left(\omega\right)$ as follows
\begin{eqnarray}
\rho(T)=\frac{\pi \Omega k_BT}{N(\varepsilon_F) \left \langle \bm{v}^2 \right \rangle}\int_{0}^{\infty} \frac{d\omega}{\omega} \frac{x^2}{sinh^2x} \alpha_{tr}^2F\left(\omega\right) \;,
\label{equ:rho}
\end{eqnarray}
where $x=\hbar\omega/(2k_BT)$ and $\Omega$ is the unit cell volume.

The electrical resistivity was calculated for the two lattice directions: for pure Zr and ZrH$_{0.5}$ with a HCP structure, the directions along primitive lattice vectors $a$ $[10\bar{1}0]$ and $c$ $[0001]$; for zirconium hydrides (at hydrogen concentration $x \geq 1$)  with a FCC or FCT structure, the directions are along primitive lattice vectors $[100]$ and close-packing direction $[111]$. We have found the relative difference between the different lattice directions is less than $2\%$. As a result, the calculated electrical resistivity presented in Sec.~\ref{sec:Results and discussion} is chosen to be the average value of all the considered directions.

\section{\label{sec:Results and discussion}Results and discussion}

\subsection{\label{sec:3b}Electron-phonon coupling and electrical resistivity of the Zr-H system}

\begin{figure}
	\includegraphics[width=\columnwidth]{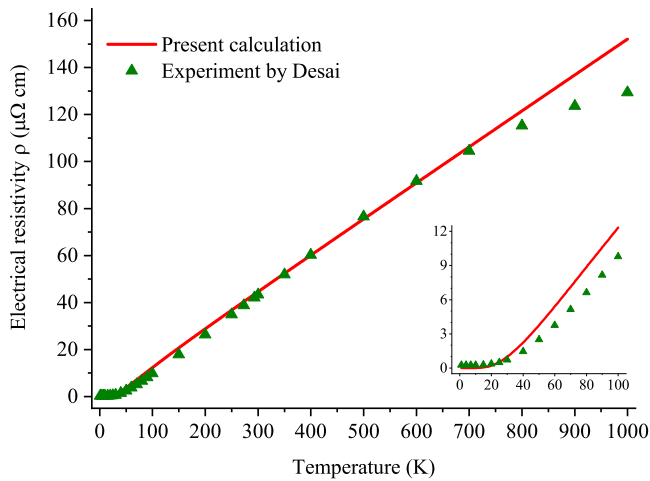}
	\caption{\label{fig:resisZr} The temperature dependence of the electrical resistivity $\rho({\rm T})$ of pure Zr.}
\end{figure}

First, we present the calculated temperature dependence of the electrical resistivity for pure Zr and the experimental results \cite{desai1984electrical} in Fig.~\ref{fig:resisZr}. It can be seen that the theoretical curve is close to the experimental data for temperature up to  $\sim$ 800 K. At a temperature above $\sim$ 800 K, significant deviations between the calculated and experimental results are observed. This situation is caused by two factors. First, at the temperature of 823 K the transformation of the $\alpha$ phase with a HCP structure into the $\beta$ phase with a body-centered cubic (BCC) structure is observed \cite{zuzek1990h}. Second, the harmonic approximation which is used in the present work to describe the EPC is incorrect at high temperature. 
As shown in the inset of Fig.~\ref{fig:resisZr}, there is a considerable difference between the calculated electrical resistivity and the experimental data at low temperature, since in this case, in addition to the electron-phonon scattering, the size effects, electron-electron scattering, scattering on impurities, etc. give significant contributions to the electrical resistance \cite{savrasov1996electron}. 
Thus, we did not investigate the electrical resistivity at a temperature below $\sim$ 200 K in this work.  It also should be mentioned that we did not consider any disorder or impurity effect in the present study, but the considered configurations in present work are stable (without imaginary phonon frequency in the spectrum).

\begin{figure}
	\includegraphics[width=\columnwidth]{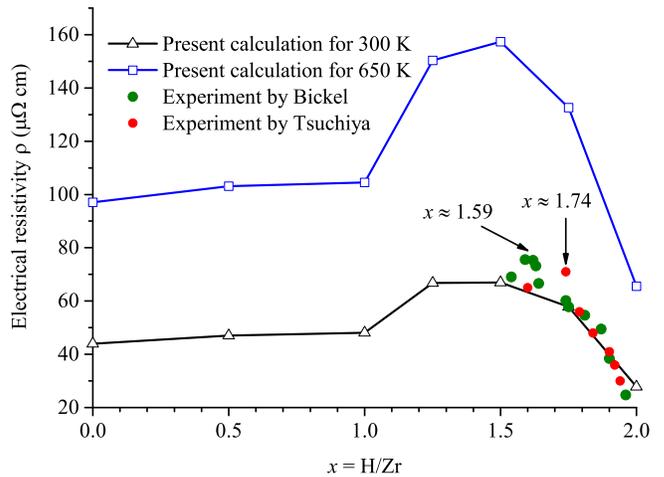}
	\caption{\label{fig:resistivityZrH} The dependence of the electrical resistivity $\rho(x)$ of the Zr-H system on a hydrogen concentration.}
\end{figure}

\begin{figure}[b]
\includegraphics[width=\columnwidth]{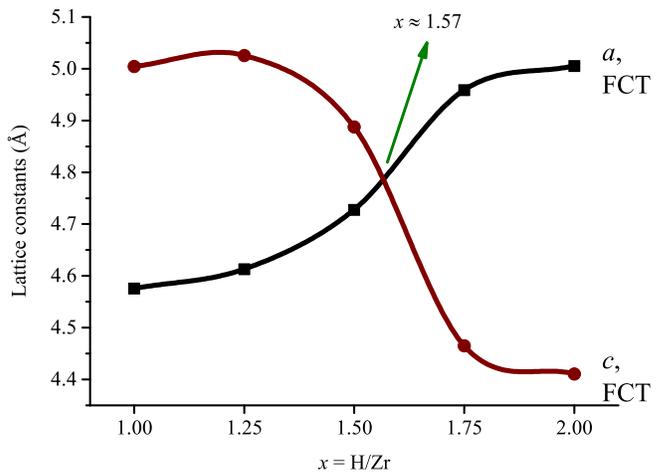}
\caption{\label{fig:LatticeConstant} The dependence of the lattice constants on the H concentration in the zirconium hydride.}
\end{figure}

\begin{figure*}
\includegraphics[width=\textwidth]{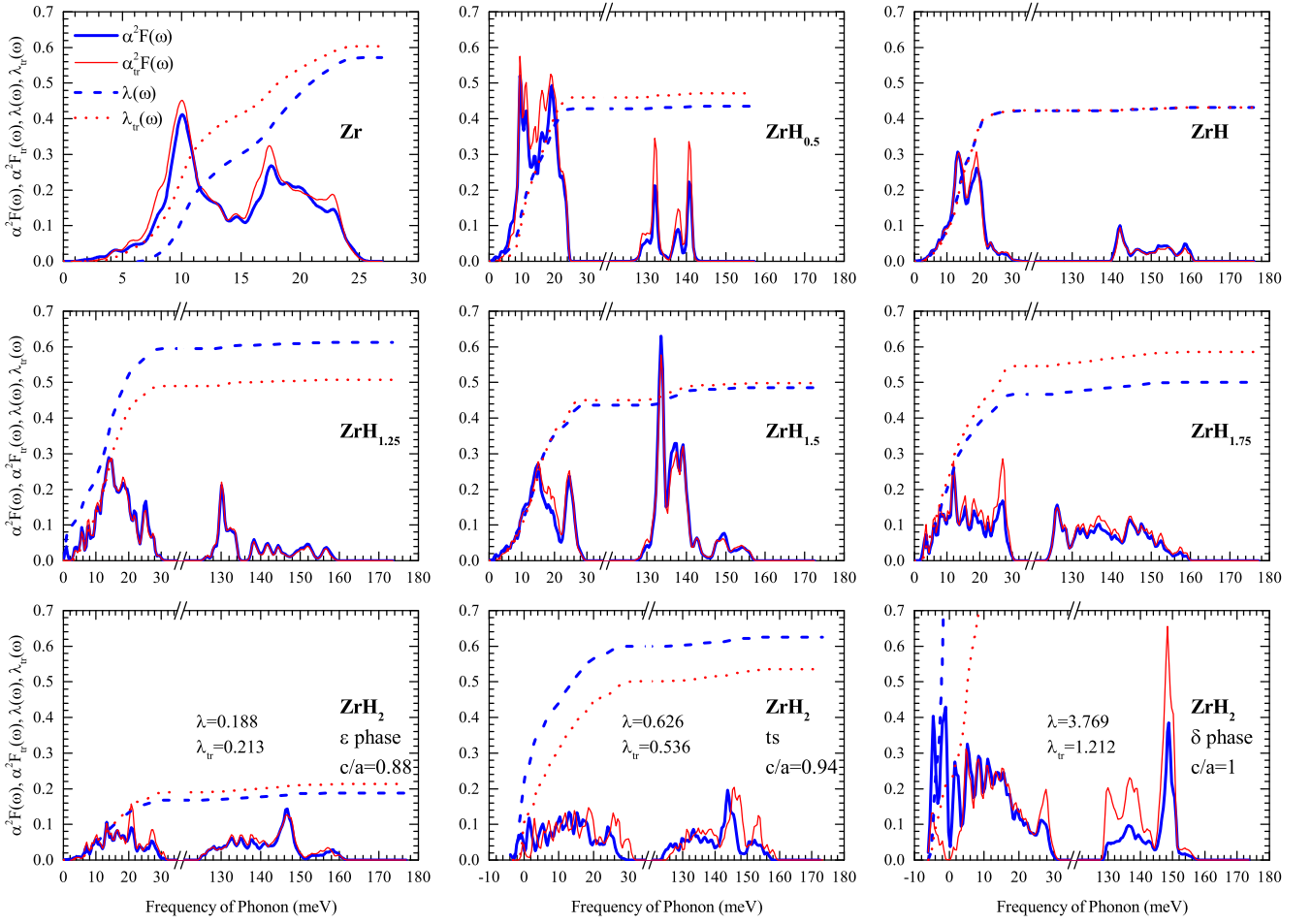}
\caption{\label{fig:Eliashberg} The calculated Eliashberg function $\alpha^2F\left(\omega\right)$, Eliashberg transport function $\alpha_{tr}^2F\left(\omega\right)$, and their strength electron-phonon coupling constant $\lambda$ and transport constant $\lambda_{tr}$ for pure zirconium, ZrH$_{0.5}$, ZrH, ZrH$_{1.25}$, ZrH$_{1.5}$, ZrH$_{1.75}$, and ZrH$_2$.}
\end{figure*}

The calculated electrical resistivity $\rho$ of the ZrH$_x$ system depending on the H concentration at 300 K is shown together with the experimental data \cite{bickel1970electrical, tsuchiya2002isotope, tsuchiya2002thermal} in Fig.~\ref{fig:resistivityZrH}. We also calculated this dependence at the typical operating temperature of the nuclear reactor (650 K). Unfortunately, in the literature there are no experimental data for this temperature. Since at a temperature of 650 K the harmonic approximation allows the correct calculation of the electrical resistance, the calculated results have practical importance. 
Figure~\ref{fig:resistivityZrH} shows that the behavior of the function $\rho(x)$ is the same at both temperature values, although at the higher temperature the function $\rho(x)$ is changed more sharply due to a stronger lattice vibration. According to both our results and the experimental results, the reduction in the electrical resistivity starts at H concentration $x_c \sim 1.5$, and zirconium hydrides become a better electrical conductor than pure Zr at H concentration $x \geqslant 1.9$ .

It should be noticed that the concentration $x_c$  is close to the concentration of the $\delta-\varepsilon$ phase transition of the Zr-H system. The $\delta$ and $\varepsilon$ phases have different values for the lattice parameter relation $c/a$. 
In Fig.~\ref{fig:LatticeConstant} , we present the dependence of the lattice parameters $c$ and $a$ on the H concentration in zirconium hydrides. It can be seen that, with the H concentration increasing the parameter $a$ is increased, while the parameter $c$ is decreased and, as a consequence, the value of the relation $c/a$ is decreased. The lattice parameters $c$ and $a$ become equal at the concentration $x \approx 1.57$, which is close to the experimental results of the $\delta-\varepsilon$ phase transition ($x \approx 1.6$) \cite{bowman1983electronic} and the concentration $x_c$. Thus, we assume that the observed behavior of the H concentration dependence of the electrical resistivity is related to the lattice tetragonal distortion in zirconium hydride. This relation will be discussed in detail in Sec.~\ref{sec:point}.

Further, we focus on the EPC in the Zr-H system. The calculated Eliashberg function $\alpha^2F\left(\omega\right)$, Eliashberg transport function $\alpha_{tr}^2F\left(\omega\right)$, and their strength electron-phonon coupling constant $\lambda\left(\omega\right)$ and transport constant $\lambda_{tr}\left(\omega\right)$ for the Zr-H system with various H concentrations are shown in Fig.~\ref{fig:Eliashberg}.
The calculation value of $\lambda=0.54$ for pure Zr indicates quite a strong electron-phonon coupling that is typical for transition metals due to the large effective mass of its $d$ electrons \cite{mott1964electrons}. Good agreement between the calculated resistivity and the experimental data allows us to conclude that the calculated spectra of $\alpha_{tr}^2F\left(\omega\right)$ are credible.

 Figure~\ref{fig:Eliashberg} shows that the Eliashberg function and Eliashberg transport function are similar in shape, and the difference between the values of $\lambda_{tr}$ and $\lambda$ does not exceed $17\%$ in all the stable structures of the ZrH$_x$ system. So we can conclude that the efficiency factor $\mu$, which gives a preferential weight from the backscattering process, gives a small contribution to the electron-phonon scattering process for the Zr-H system \cite{savrasov1996electron}.
It should be noticed that the difference between $\lambda_{tr}$ and $\lambda$ is very small in ZrH, ZrH$_{1.5}$, and ZrH$_2$. In the cases of ZrH$_{1.25}$ and ZrH$_{1.75}$, which were not observed in the experiment, this difference is observable, which indicates a Fermi surface nesting which is related to the instability of these systems. Therefore, we will discuss below only ZrH, ZrH$_{1.5}$, and ZrH$_2$, which correspond to the experimentally observed $\gamma$, $\delta$, and $\varepsilon$ zirconium hydride phases.
 In addition, it can be seen from the case of the unstable structures of ZrH$_2$ with $c/a = 0.94$ and $c/a = 1$ that the electron backscattering weakens the strong coupling. In particular, the system is not able to hold a strong EPC since strong coupling leads to the instability of the system. As a result, the system transforms into a new structure with weaker coupling. It seems that there is a relevance between the electron backscattering and the phase transition in the Zr-H system. Especially, there is a significant difference between $\alpha^2F\left(\omega\right)$ and $\alpha_{tr}^2F\left(\omega\right)$ in $\delta$-ZrH$_2$ due to the electron backscattering, which may cause a strong Fermi surface nesting \cite{crespi1992possible} and a phase transition. It is clearly seen for ZrH$_2$ that the decrease in the value of  $c/a$ from 1 to 0.88 leads to the increase of the relation $\lambda_{tr}/\lambda$ from 0.32 to 1.14. This means that the EPC becomes weaker due to the tetragonal distortion, and the strong Fermi surface nesting of the $\delta$ phase recedes. The weak EPC in $\varepsilon$-ZrH$_2$ can explain why ZrH$_2$ is a better conductor than pure Zr.

The contribution of H atoms to the electron-phonon scattering is also an interesting issue. It was found that the H-character phonons give a small contribution to $\lambda$ and $\lambda_{tr}$ in the Zr-H system as shown Fig.~\ref{fig:Eliashberg}. In particular, the contribution of the H-character phonon to $\lambda$ is $2\%$, $10\%$, and $11\%$ in ZrH, ZrH$_{1.5}$, and ZrH$_2$, respectively. The small contribution of H atoms can be explained by the presence of strong Zr-H chemical bonds \cite{quijano2009electronic}, which limits the H vibration modes on a small scale. This fact can also explain why the harmonic approximation correctly describes the phonon structure in the Zr-H system. The contribution of the H-character phonon modes to EPC increases with the increasing of the H concentration.

\subsection{\label{sec:point}Electron-phonon-driven phase transition and reduction of electrical resistivity}

\begin{figure}
\includegraphics[width=\columnwidth]{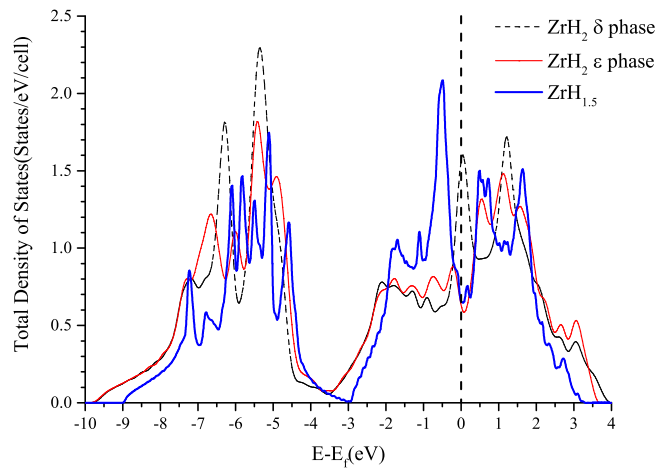}
\caption{\label{fig:EDOS} The total EDOS in ZrH$_2$ $\delta$ and $\varepsilon$ phases and ZrH$_{1.5}$. The Fermi level is set to the zero point.}
\end{figure}

\begin{figure}
\includegraphics[width=\columnwidth]{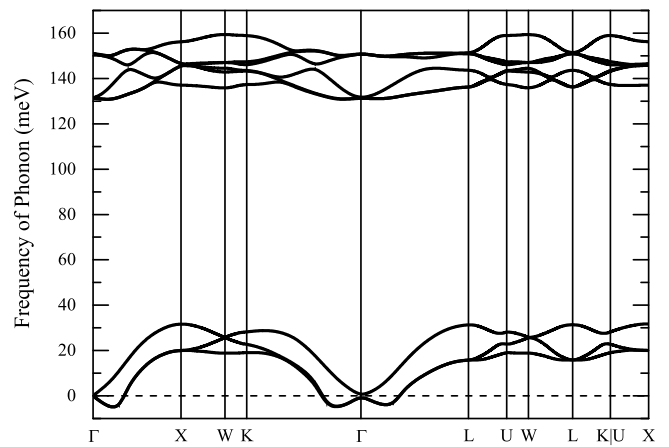}
\caption{\label{fig:phononZrH2} The phonon spectrum of the ZrH$_2$ $\delta$ phase. }
\end{figure}

\begin{figure}
\includegraphics[width=\columnwidth]{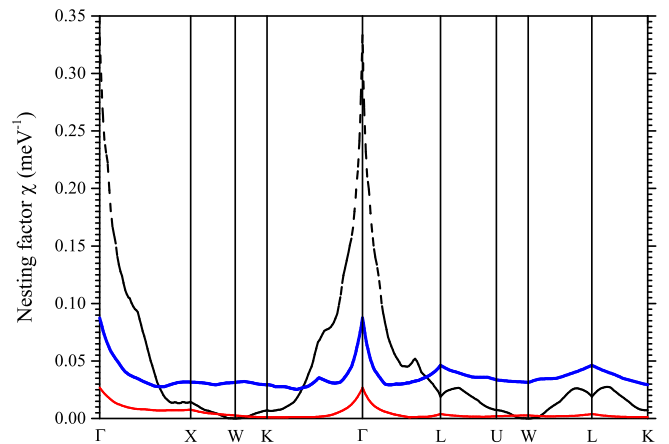}
\caption{\label{fig:nesting} The Fermi surface nesting factor for the ZrH$_2$ $\delta$ and $\varepsilon$ phases, and ZrH$_{1.5}$. The black dashed line is the Fermi surface nesting factor of the ZrH$_2$ $\delta$ phase, the red solid line is the Fermi surface nesting factor of the ZrH$_2$ $\varepsilon$ phase, and the blue solid line represents the Fermi surface nesting factor of ZrH$_{1.5}$.}
\end{figure}

In this section we will discuss in detail the influence of the tetragonal distortion of the FCC structure on the electrical properties of ZrH$_2$. 
As we said, the instability of the FCC structure plays an important role in the reduction of the electrical resistivity; it leads to the $\delta-\varepsilon$ phase transition and also weakens the electron-phonon coupling in the Zr-H system.  There are three indications of the structural instability of the $\delta$ phase: the high peak of the EDOS at the Fermi level, imaginary phonon frequencies and the extremely high value of the Fermi surface nesting factor, and they correspond to the electron-, phonon-, and electron-phonon- driven mechanisms of the tetragonal distortion in the Zr-H system.
There are two main electron-driven mechanisms, the first is the splitting of the bands at the Fermi level in the $\Gamma-L$ direction due to Jahn-Teller effect \cite{cantrell1984x, wolf2000first, miwa2002first, quijano2009electronic}, and the second is the reduction of $N(\varepsilon_F)$ by a shift in energy of the band along the $\Gamma-K$ direction \cite{kul1999electronic, quijano2009electronic}. The key point is that the two electron-driven mechanisms lead to the reduction in $N(\varepsilon_F)$ from the $\delta$ phase to the $\varepsilon$ phase (Fig.~\ref{fig:EDOS}). We can see that the high $N(\varepsilon_F)$ splits into two peaks near the Fermi level through the $\delta-\varepsilon$ phase transition, and the system obtains structural stability in this way. The phonon spectrum also shows the structural instability of the ZrH$_2$ $\delta$ phase. As shown in Fig.~\ref{fig:phononZrH2}, in the $\Gamma-K$, $\Gamma-X$, and $\Gamma-L$ directions the Zr-character phonon modes have imaginary frequencies around the $\Gamma$ point. As a consequence, in the $\Gamma-K$ and $\Gamma-L$ directions the signs of electron-driven mechanisms of phase transition were observed, so we can assume that the phase transition is related to the interaction between electrons and phonons.

According to Eq.~\ref{equ:rho}, there are three parameters which influence the electrical resistivity: the conduction electron concentration $n(\varepsilon_F)=N(\varepsilon_F)/\Omega$, the transport constant $\lambda_{tr}$ and the average square of the Fermi velocity $\left \langle \bm{v}^2 \right \rangle$. In Fig.~\ref{fig:EDOS} we also compare the EDOS between the ZrH$_2$ $\varepsilon$ phase and ZrH$_{1.5}$ with the FCC(T) structure since ZrH$_{1.5}$ can be considered the energetically stable configuration of the $\delta$ phase (the ZrH$_2$ $\delta$ phase is unstable). We can see that there is no significant difference in the EDOS at the Fermi level between the ZrH$_2$ $\varepsilon$ phase and ZrH$_{1.5}$. In fact, the value of the transport constant $\lambda_{tr}$ of 0.498 for ZrH$_{1.5}$ is about 2.338 times higher than the value of 0.213 for ZrH$_2$, and this is in good agreement with the value of 2.4 for the ratio of $\rho$(ZrH$_{1.5}$)/$\rho$(ZrH$_2$). Thus, we conclude that the reduction in the electrical resistivity is defined by the change in $\lambda_{tr}$; the influences of the conduction electron concentration and the Fermi velocity are small, and they equalize each other.

\begin{figure}
\includegraphics[width=\columnwidth]{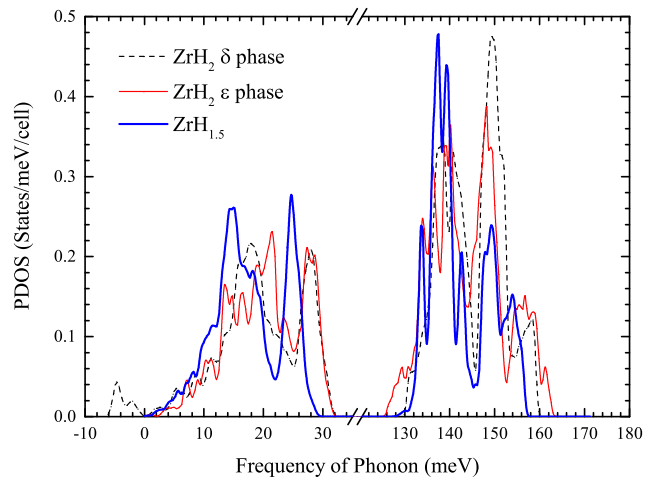}
\caption{\label{fig:PDOS} The PDOS of the ZrH$_2$ $\delta$, $\varepsilon$ phases and ZrH$_{1.5}$. }
\end{figure}

Two factors influence $\lambda_{tr}$: the Fermi surface nesting and the electron-phonon coupling matrix elements. To evaluate the Fermi surface nesting, we calculated the distribution of the nesting factor
$\chi(\bm{q})=\sum_{\substack{\\ ij\bm{k}}}{\delta(\varepsilon_{j\bm{k}}-\varepsilon_F)\delta(\varepsilon_{i\bm{k}+\bm{q}}-\varepsilon_F)}$
in the reciprocal space, as shown in Fig.~\ref{fig:nesting}. As we said, the strong nesting is the electron-phonon-driven mechanism of the $\delta-\varepsilon$ phase transition of the Zr-H system. 
It can clearly be seen that in the ZrH$_2$ $\delta$ phase, the nesting is strong around the $\Gamma$ point in the $\Gamma-K$, $\Gamma-X$, $\Gamma-L$ directions where the imaginary phonon frequencies appear. It can also be found that the nesting factor has its maximum at the $\Gamma$ point. It is interesting that the $\Gamma$ point is like a singular point which is the center of imaginary phonon frequencies, but the Zr-character acoustic phonon frequencies at $\Gamma$ point are close to 0. The Zr-character acoustic phonon modes are following the ``2+1" splitting \cite{giannozzi1991ab}, the frequencies of the two degenerate modes are imaginary, and the frequencies of split mode are positive. This may be caused by the fact that the nesting vector (points to the maximum of the nesting factor) $\bm{q} \to 0$ ($\Gamma$ point) and the $\Gamma$ point as the point with the highest degeneracy cannot be modified, so the period-lattice distortion and the charge-density modulation have not been observed in ZrH$_2$, unfortunately. But the Fermi surface nesting behavior in ZrH$_2$ can still help us to understand the modification of the electron and phonon structures of the system, especially the changes around the $\Gamma$ point. In particular, through the $\delta-\varepsilon$ phase transition, the nesting factor $\chi$ not only strongly decreases around the $\Gamma$ point but also has a global reduction. If we compare the stable ZrH$_{1.5}$ and ZrH$_2$, we can see that the nesting in ZrH$_{1.5}$ is stronger than in ZrH$_2$. 
In particular, we can see that in the system the effect of the electron-phonon coupling on the electron transport is weakened due to the modification of the electronic structure, which leads to the decrease in the electrical resistance. 
Also, such a system with strong Fermi surface nesting is unstable, and it may lead to a phase transition.

\begin{figure}[b]
	\includegraphics[width=\columnwidth]{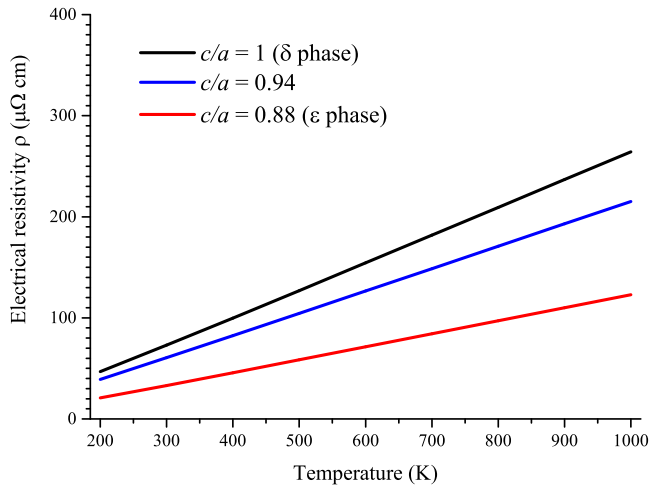} 
	\caption{\label{fig:resistivityZrH2}The temperature dependence of the electrical resistivity in the ZrH$_2$ $\delta$ and $\varepsilon$ phases and the ts.}
\end{figure}

It is also important to investigate the influence of the electron-phonon coupling matrix elements, which indicate the strength of the single-scattering process. To achieve this goal, we calculated the changes in the phonon density of states (PDOS) through the $\delta-\varepsilon$ transition, as shown in Fig.~\ref{fig:PDOS}. We can see that the frequency range of the H-character optical phonon in both the ZrH$_2$ $\delta$ phase and ZrH$_{1.5}$ is narrower than in the ZrH$_2$ $\varepsilon$ phase, and the gap between the acoustic and optical phonons in both the ZrH$_2$ $\delta$ phase and ZrH$_{1.5}$ is higher than in the ZrH$_2$ $\varepsilon$ phase. Under a high-symmetry condition (cubic structure), phonon modes in the system have a higher degeneracy, which leads to the high peak and narrow range of the H-character optical phonon frequency, and this might be a reason why the H-character optical phonon plays a more important role in the $\delta$ phase.
The shapes of the PDOS and $\alpha^2F\left(\omega\right)$ are similar, in fact, $\alpha^2F\left(\omega\right)$ can also be seen as a modification of the PDOS with a weight from the electron-phonon coupling. Figure~\ref{fig:PDOS} shows that the PDOSs in the $\delta$ and $\varepsilon$ phases have no significant difference, but the comparison of $\alpha^2F\left(\omega\right)$ between the $\delta$ and $\varepsilon$ phases in Fig.~\ref{fig:Eliashberg} shows that the electron-phonon coupling in the $\delta$ phase is much stronger than in the $\varepsilon$ phase. Thus, we can conclude that the electron-phonon coupling matrix elements play the defining role in the scattering process.

In addition, to clarify that the tetragonal distortion of a FCC structure is the reason for the decrease in the electrical resistance, we consider an imaginary transition state (ts) of ZrH$_2$ with the lattice volume and ratio $c/a$ being the average of the $\delta$ and $\varepsilon$ phases. The calculated results of the ts show that this state has a structural instability weaker than the $\delta$ phase: the values of $\lambda=0.626$ and $\lambda_{tr}=0.536$ are lower than in the case of the $\delta$ phase, and higher than in the case of the $\varepsilon$ phase.
In the end, we show the calculated results of $\rho(T)$ for the ZrH$_2$ $\delta$, $\varepsilon$ phases and the ts in Fig.~\ref{fig:resistivityZrH2}. We can clearly see that the electrical resistivity of ZrH$_2$ decreases through the tetragonal distortion. In the $\delta$ phase the electrical resistivity is much higher, and the temperature dependence $\rho(T)$ is steeper than in the $\varepsilon$ phase. It should be noticed that the results for the $\delta$ phase and ts are not physical because of the imaginary phonon frequencies. 
We can conclude that the high electrical resistivity in the $\delta$ phase is caused by its strong electron-phonon coupling, and it comes from the structural instability of the FCC structure. We can see that through the tetragonal distortion the imaginary phonon frequency and the strong electron-phonon coupling are eliminated. As a result, the stable state of the $\varepsilon$ phase with weak electron-phonon coupling strength is obtained, and it causes the reduction in the electrical resistivity of the ZrH$_x$ system at a high hydrogen concentration ($x>1.5$). In fact, the strong electron-phonon coupling in the $\delta$ phase can also be seen as a result of its high $N_{F}$. The high $N_{F}$ provides a large number of backscattering electrons and gives a contribution to the strong Fermi surface nesting. Thus, we can conclude that the strong electron-phonon coupling has significant relevance to the structural instability of the $\delta$ phase.

\section{\label{sec:Summary and conclusions}Summary and conclusions}
In the present paper, the nature of the anomalous decrease in the electrical resistivity of the Zr-H system with the increase of hydrogen concentration (at H/Zr$>$1.5) was studied within the framework of the electron density functional theory and the density functional perturbation theory, using the optimized norm-conserving Vanderbilt pseudopotential method. To understand the nature of the reduction, a comprehensive study of the main factors decreasing the resistance was carried out.

It was shown that the electrical resistivity in the Zr-H system is not determinated by the conduction electron concentration but the change in electron-phonon coupling. 
We analyzed the electron-phonon coupling in the Zr-H system by means of the Eliashberg function $\alpha^2F\left(\omega\right)$ and the Eliashberg transport function $\alpha_{tr}^2F\left(\omega\right)$ in detail.
It was shown that with the increasing of H concentration in Zr, the change in the electron-phonon coupling strength is not great except in ZrH$_2$. It was also found that in the stable configuration of the Zr-H system the difference between $\alpha^2F\left(\omega\right)$ and $\alpha_{tr}^2F\left(\omega\right)$ is small, but it is large in the ZrH$_2$ $\delta$ phase, which is unstable. It was established that the significant difference between $\alpha^2F\left(\omega\right)$ and $\alpha_{tr}^2F\left(\omega\right)$ in the ZrH$_2$ $\delta$ phase is caused by the backscattering of electrons due to the strong Fermi surface nesting, which is one of the reasons why ZrH$_2$ transforms from the $\delta$ phase into the $\varepsilon$ phase.
It was also found that the strongly reduced Fermi surface nesting due to the $\delta-\varepsilon$ phase transion is one main factor decreasing the resistance of Zr-H system.
From this, the correlation between the $\delta-\varepsilon$ phase transition and the reduction of electrical resistivity was clarfied. 
Another main factor decreasing the resistance is the strong reduction of the electron-phonon coupling matrix elements, which indicate the strength of single electron-phonon scattering process.
It was shown that the ZrH$_2$ $\delta$ phase has a resistivity significantly larger than that of the $\varepsilon$ phase due to the strong electron-phonon coupling, which leads to the structural instability of the $\delta$ phase. The tetragonal lattice distortion due to the $\delta-\varepsilon$ phase transition of ZrH$_2$ eliminates imaginary phonon frequencies and the strong electron-phonon coupling.

\begin{table*}
	\caption{\label{tab:table1}The lattice parameters ($a$ and $c$) for the calculated ZrH$_x$ systems with H atoms located all at T or all at O sites.}
	\renewcommand\arraystretch{1.2}
	\begin{ruledtabular}
		\begin{tabular}{ccccccc}
			\\ [-2 ex]
			System&Structure&$a$ $(\rm{\AA})$&$c$ $(\rm{\AA})$&$c/a$&Method&Reference\\ [-2 ex]
			\\ \hline
			\\[-2.5 ex]
			Pure zirconium&HCP&3.2346&5.1678&1.5976&GGA&Present work\\
			& &3.2317&5.1476&1.5928&Exp. (298K)&\onlinecite{zuzek1990h}\\
			& &3.213&5.159&1.605&GGA&\onlinecite{wang2012first}\\
			& &3.23&5.18&1.60&GGA&\onlinecite{domain2002atomic}\\
			Zr$_2$H&HCP(T), C1&3.2603&5.3962&1.6551&GGA&Present work\\
			& &3.3&5.145&1.56&Exp. (293K)&\onlinecite{zhao2008identification}
			\\
			& &3.313&5.549&1.675&GGA&\onlinecite{wang2012first}\\
			&HCP(T), C2&3.2433&5.4892&1.6925&GGA&Present work\\
			& &3.26&5.447&1.671&GGA&\onlinecite{wang2012first}\\
			&FCT(T), C3&4.7597&4.4595&0.9369&GGA&Present work\\
			& &4.676& & &GGA&\onlinecite{wang2012first}\\
			&FCT(O), C4&4.4334&4.8431&1.0924&GGA&Present work\\
			& &4.575& & &GGA&\onlinecite{wang2012first}\\
			ZrH&FCT(T)&4.5754&5.0045&1.0938&GGA&Present work\\
			& &4.5957&4.9686&1.081&Exp. (293K)&\onlinecite{mueller2013metal}\\
			& &4.61&5.04&1.093&GGA&\onlinecite{wang2012mechanical}\\
			ZrH$_{1.25}$&FCT(T)&4.6130&5.0252&1.0894&GGA&Present work\\
			& &4.79&5.20&1.086&GGA&\onlinecite{wang2012mechanical}\\
			ZrH$_{1.5}$&FCT(T)&4.7276&4.8874&1.0338&GGA&Present work\\
			& &4.65&4.96&1.067&Exp. (320K)&\onlinecite{bowman1983electronic}\\
			& &4.62&4.83&1.046&GGA&\onlinecite{wang2012mechanical}\\
			ZrH$_{1.75}$&FCT(T)&4.9585&4.4651&0.9005&GGA&Present work\\
			& &4.9087&4.5220&0.9212&Exp. (118K)&\onlinecite{cantrell1984x}\\
			& &4.97&4.47&0.899&GGA&\onlinecite{wang2012mechanical}\\
			ZrH$_2$&FCT(T)&5.0053&4.4106&0.8812&GGA&Present work\\
			& &4.9808&4.4336&0.8901&Exp. (108K)&\onlinecite{cantrell1984x}\\
			& &4.975&4.447&0.894&Exp. (294K)&\onlinecite{niedzwiedz199391zr}\\
			& &4.982&4.449&0.893&Exp. (320K)&\onlinecite{bowman1983electronic}\\
			& &5.021&4.432&0.883&GGA&\onlinecite{quijano2009electronic}\\
			&FCC(T)&4.8089& & &GGA&Present work\\
			& &4.817& & &GGA&\onlinecite{quijano2009electronic}\\
			& &4.82& & &GGA&\onlinecite{domain2002atomic}\\
			& &4.804& & &LDA&\onlinecite{wolf2000first}\\
			\\[-2.5 ex]
		\end{tabular}
	\end{ruledtabular}
\end{table*}

\begin{figure*}
	\subfigure[ C1 ] 
	{
		\begin{minipage}[b]{0.2\textwidth}
			\includegraphics[height=5cm]{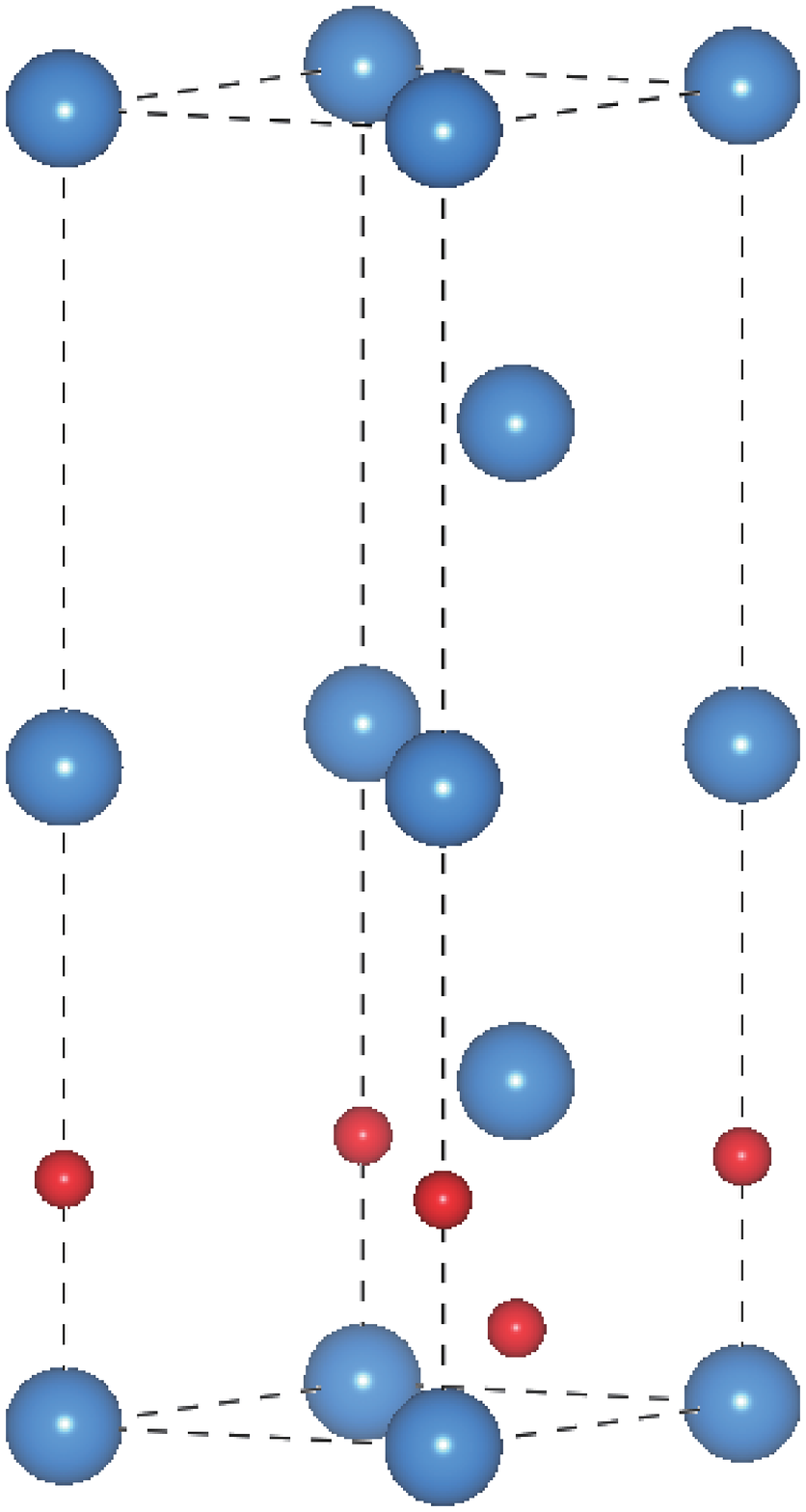} 
		\end{minipage}
	}
	\subfigure[ C2 ]
	{
		\begin{minipage}[b]{0.2\textwidth}
			\includegraphics[height=5cm]{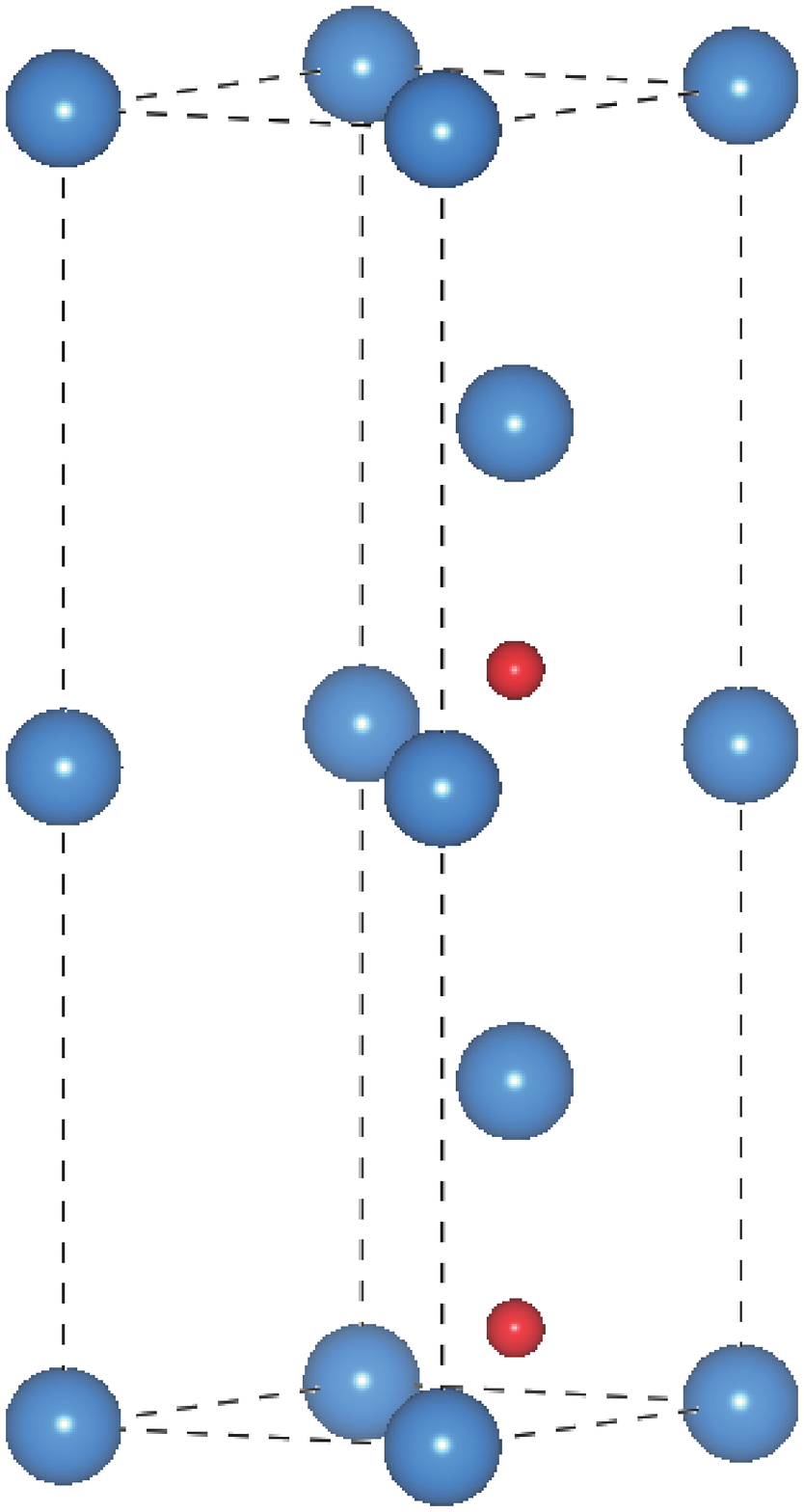} 
		\end{minipage}
	}
	\subfigure[ C3 ]
	{
		\begin{minipage}[b]{0.2\textwidth}
			\includegraphics[height=2.6cm]{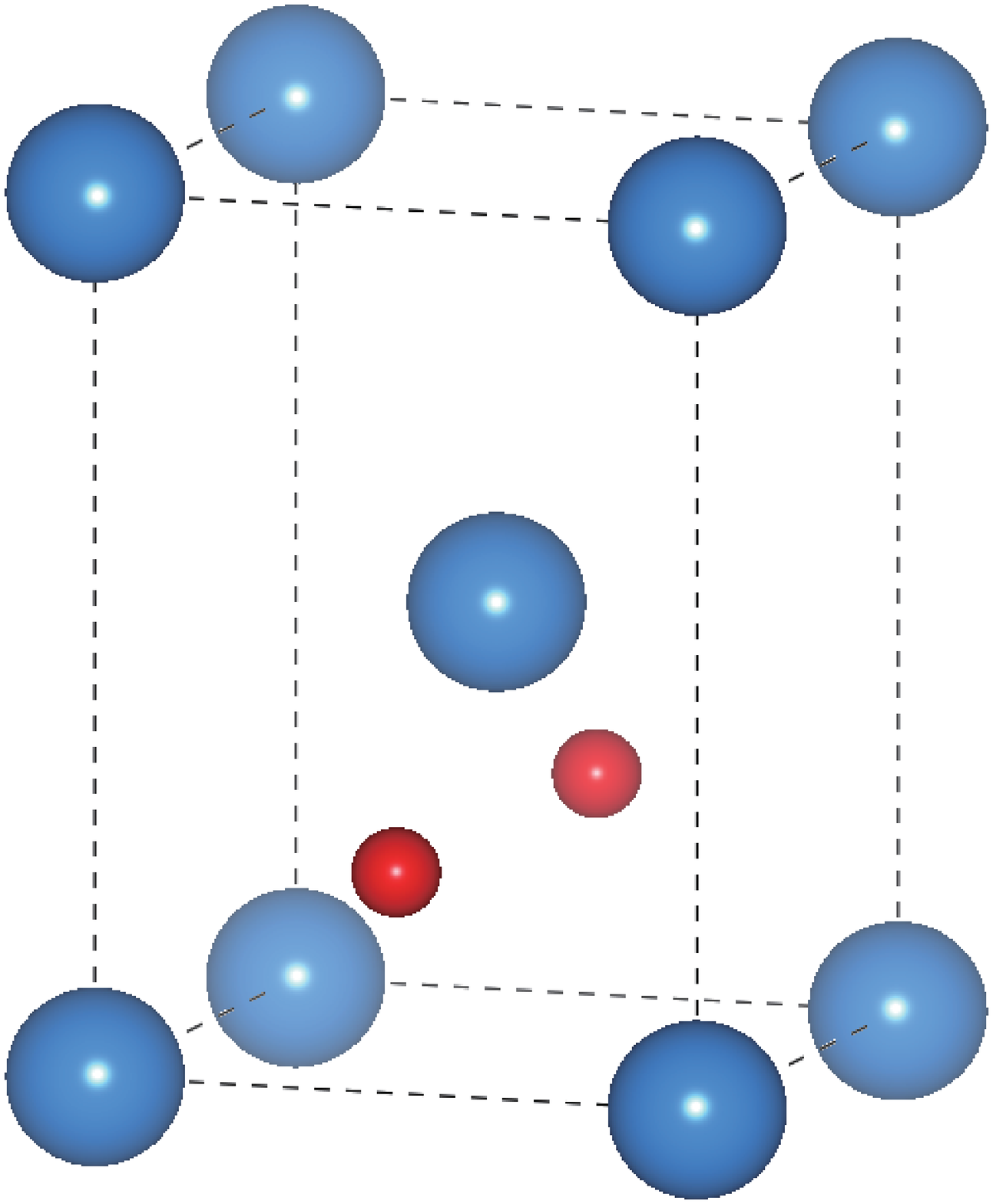} 
		\end{minipage}
	}
	\subfigure[ C4 ]
	{
		\begin{minipage}[b]{0.2\textwidth}
			\includegraphics[height=2.6cm]{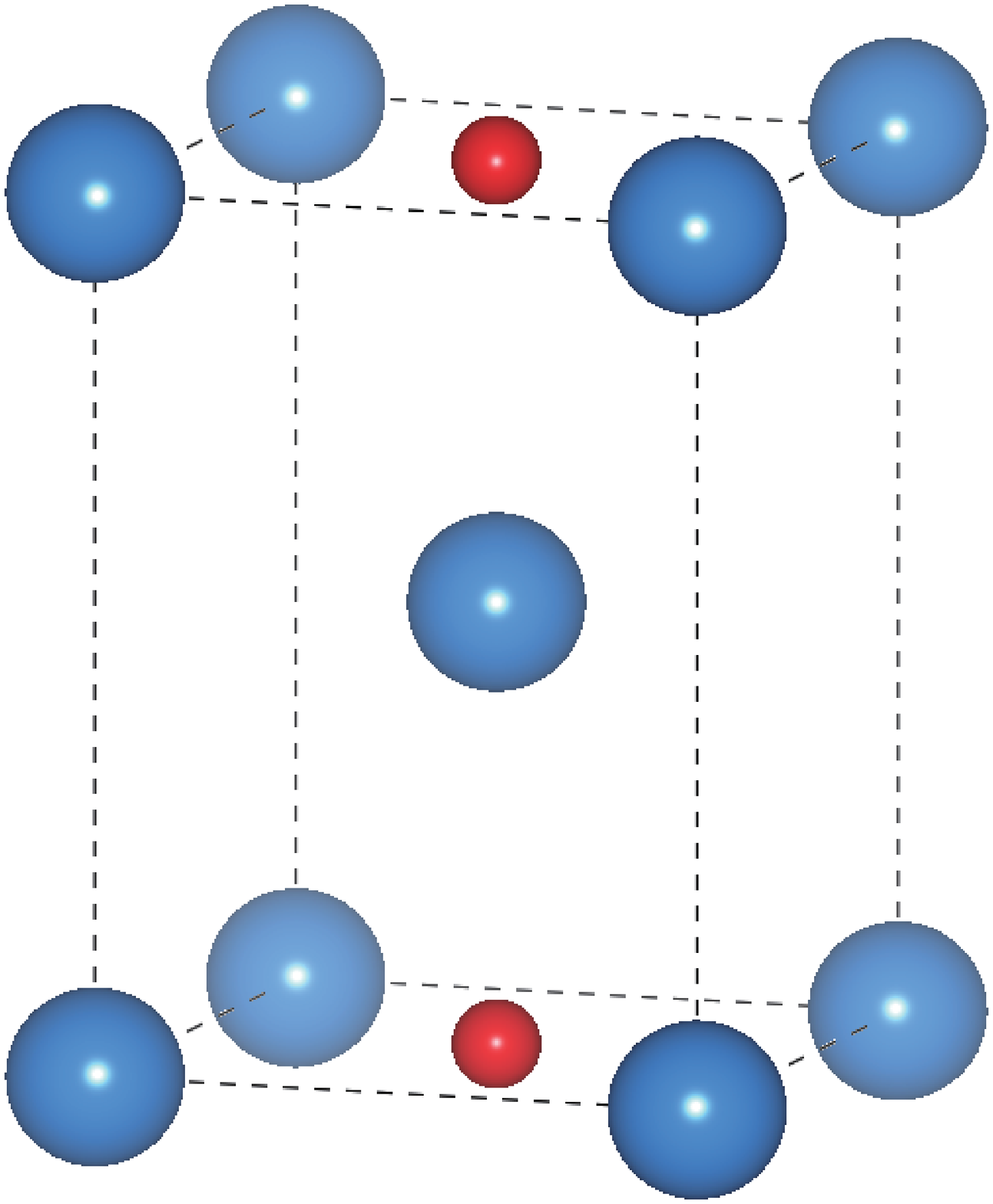} 
		\end{minipage}
	}
	\caption{\label{fig:C}The four configurations of ZrH$_{0.5}$ which were considered in our calculations.}
\end{figure*}

\begin{figure}[b]
	\includegraphics[width=\columnwidth]{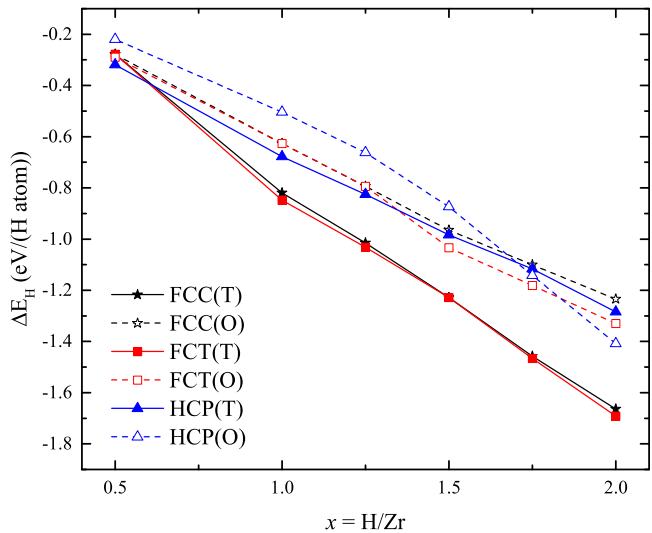}
	\caption{\label{fig:energy}The dissolution energy of hydrogen in zirconium as a function of the H concentration.}
\end{figure}

\begin{figure}
	\subfigure[HCP(T), C1]
	{
		\begin{minipage}[b]{\columnwidth}
			\includegraphics{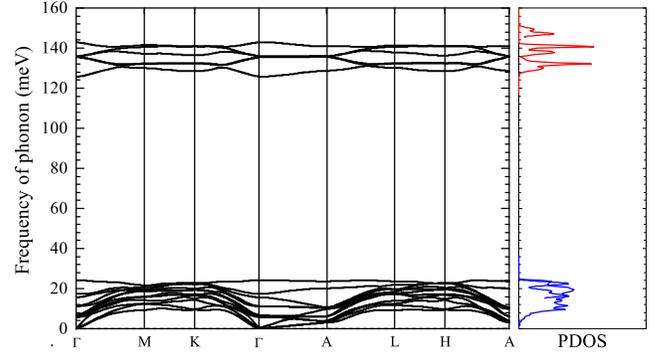} \\
		\end{minipage}
	}
	\subfigure[HCP(T), C2]
	{
		\begin{minipage}[b]{\columnwidth}
			\includegraphics{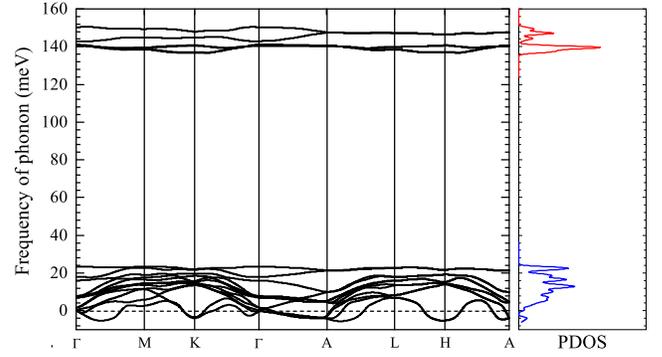} \\
		\end{minipage}
	}
	\subfigure[FCT(T), C3]
	{
		\begin{minipage}[b]{\columnwidth}
			\includegraphics{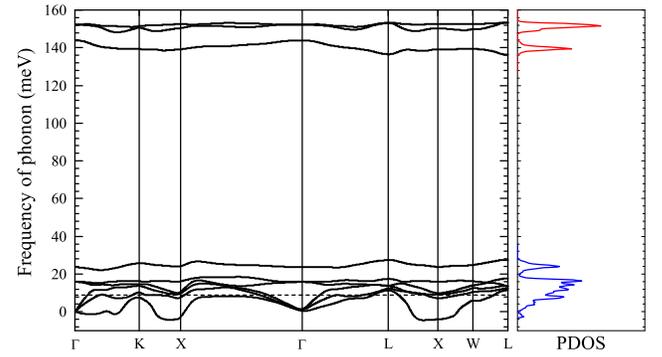} \\
		\end{minipage}
	}
	\subfigure[FCT(O), C4]
	{
		\begin{minipage}[b]{\columnwidth}
			\includegraphics{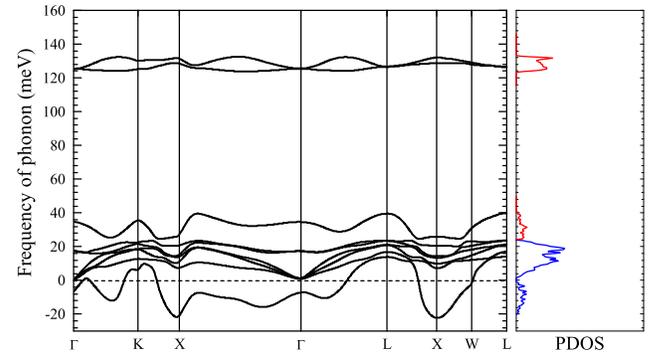} \\
		\end{minipage}
	}
	\caption{\label{fig:phononZr2H}The phonon spectra for Zr$_2$H having low symmetry configurations C1 and C2 with HCP(T) structure and high symmetry configurations C3 and C4 with FCT(T), and FCT(O) structures, respectively.}
\end{figure}

\begin{figure}
	\includegraphics[width=\columnwidth]{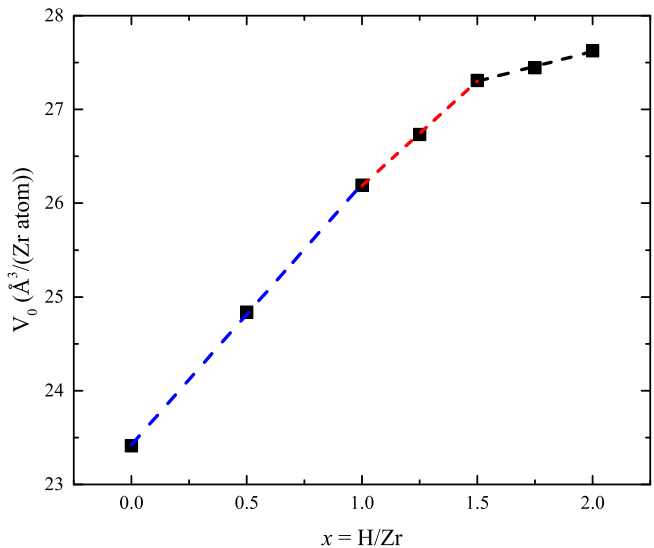}
	\caption{\label{fig:volume}The unit cell volume of the Zr-H system as a funciton of the hydrogen concentration. Here the three dash lines are the results of linear fitting in the ranges of $0\leqslant x\leqslant 1$, $1\leqslant x\leqslant 1.5$ and $1.5\leqslant x\leqslant 2$.}
\end{figure}

\begin{acknowledgements}
The research was carried out within the framework of the grant of the Program for Enhancing Competitiveness of Tomsk Polytechnic University.
\end{acknowledgements}

\appendix*

\section{Structure of Zr-H system} \label{APP}

Before the calculation of the electron-phonon coupling in the Zr-H system, we determinated the lattice structure of this system. 
The lattice structure of Zr-H systems has been well studied in both experimental and theoretical studies \cite{zuzek1990h, narang1977location, khoda1982determination, bowman1983electronic, niedzwiedz199391zr, ackland1998embrittlement, zhao2008identification, wang2012first, wang2012mechanical}. In the present work, for zirconium hydrides (with hydrogen concentration $x\geq1$), we studied the stable configurations which were defined in Ref.~\cite{wang2012mechanical}.
For ZrH$_{0.5}$, according to the results of previous works, there is an ambiguity.
The metastable state of ZrH$_{0.5}$ with a HCP structure with low-symmetry was identified by transmission electron microscopy and first-principles calculations \cite{zhao2008identification}. However, it was reported in Ref.~\cite{wang2012first} that ZrH$_{0.5}$ has a FCC structure in a high-symmetry configuration.  Hence, for ZrH$_{0.5}$ we considered four configurations, C1-C4, to study its structural stability, as shown in Fig.~\ref{fig:C}. 

First, we calculated the dissolution energy $\Delta E_{\rm{H}}$ of hydrogen in zirconium
\begin{eqnarray}
\Delta E_{\rm{H}}=\frac{E_{{\rm{Zr}}_a{\rm{H}}_b}-aE_{\rm{Zr}}-\frac{b}{2}E_{\rm{H_2}}}{b}\;,
\end{eqnarray}
where $E_{{\rm{Zr}}_a{\rm{H}}_b}$, $E_{\rm{Zr}}$, and $E_{\rm{H_2}}$ are the total energies of Zr$_a$H$_b$, pure HCP Zr (ground state) and a H$_2$ molecule, respectively. 
$\Delta E_{\rm{H}}$ of the various ZrH$_x$ systems as a function of H concentration are shown in Fig.~\ref{fig:energy}. It can be seen from Fig.~\ref{fig:energy} that it is energetically most advantageous for the ZrH$_x$ system to have the HCP(T) lattice structure at $x = 0.5$ and the FCT(T) structure at $1\leqslant x \leqslant 2$. At a concentration of $1.5\leqslant x \leqslant 1.75$, the dissolution energy of H in Zr with the FCT(T) structure exceeds the dissolution energy in Zr with the FCC(T) structure by an amount not exceeding 0.1 eV/H atom. As a result, under external perturbations (for example, radiation or a high temperature), the ZrH$_x$ system can pass from the FCT structure to the FCC structure. It can also be seen that the H concentration dependence of the dissolution energy $\Delta E_{\rm{H}}$ is linear, particularly with a slope coefficient of $-0.900 \pm 0.026$ eV/H atom. 
It should be pointed out that configuration C1 of ZrH$_{0.5}$, which is presented as the HCP(T) structure in Fig.~\ref{fig:energy}, has the lowest total energy; however, this energy is close to the total energies of configurations C3 and C4 with the FCT lattice (the difference in the total energy is less than 0.4 meV/Zr atom).
Since the energy difference between the HCP and FCT structures of ZrH$_{0.5}$ is too small to determinate the stable structure, the phonon spectra of ZrH$_{0.5}$ with configurations C1-C4 were analyzed to study its structural stability (see Fig.~\ref{fig:phononZr2H}). It can clearly be seen that the H location in Zr is important to the structural stability of the Zr-H system: configurations C2-C4 have the imaginary phonon frequency, while configuration C1 does not.

The calculated lattice parameters ($a$ and $c$) of pure Zr and the ZrH$_x$ systems are shown in Table~\ref{tab:table1} and have very good agreement with the previous calculation results and experimental data \cite{zuzek1990h, wang2012first, domain2002atomic, zhao2008identification, mueller2013metal, bowman1983electronic, cantrell1984x, niedzwiedz199391zr, quijano2009electronic}. It is observed that the ratio $c/a$ decreases with increasing H concentration in Zr, as in experiments \cite{bowman1983electronic, cantrell1984x}. As the intrinsic electrical resistivity depends on the crystal volume [Eq.~\ref{equ:rho}],  it is interesting to study a change in the crystal volume of the Zr-H system depending on the hydrogen concentration (see Fig.~\ref{fig:volume}). This dependence has a linear character with different slopes at hydrogen concentration ranges of $0 \leqslant x\leqslant 1$, $1 \leqslant x\leqslant 1.5$ and $1.5 \leqslant x\leqslant 2$. 
The linear approximation by the least-squares method gives lines with slope coefficients of $2.78 \pm 0.04$ $\rm{\AA}^3$/Zr atom in the range of $0 \leqslant x\leqslant 1$, $2.23 \pm 0.04$ $\rm{\AA}^3$/Zr atom in the range of $1 \leqslant x\leqslant 1.5$ and $0.63 \pm 0.05$ $\rm{\AA}^3$/Zr atom in the range of $1.5 \leqslant x\leqslant 2$. The change in the slope in the dependence of the crystal volume on the hydrogen concentration is due to the phase transition in the Zr-H system.

\bibliographystyle{apsrev4-1}
\bibliography{apssamp}

\begin{thebibliography}{47}%
\makeatletter
\providecommand \@ifxundefined [1]{%
 \@ifx{#1\undefined}
}%
\providecommand \@ifnum [1]{%
 \ifnum #1\expandafter \@firstoftwo
 \else \expandafter \@secondoftwo
 \fi
}%
\providecommand \@ifx [1]{%
 \ifx #1\expandafter \@firstoftwo
 \else \expandafter \@secondoftwo
 \fi
}%
\providecommand \natexlab [1]{#1}%
\providecommand \enquote  [1]{``#1''}%
\providecommand \bibnamefont  [1]{#1}%
\providecommand \bibfnamefont [1]{#1}%
\providecommand \citenamefont [1]{#1}%
\providecommand \href@noop [0]{\@secondoftwo}%
\providecommand \href [0]{\begingroup \@sanitize@url \@href}%
\providecommand \@href[1]{\@@startlink{#1}\@@href}%
\providecommand \@@href[1]{\endgroup#1\@@endlink}%
\providecommand \@sanitize@url [0]{\catcode `\\12\catcode `\$12\catcode
  `\&12\catcode `\#12\catcode `\^12\catcode `\_12\catcode `\%12\relax}%
\providecommand \@@startlink[1]{}%
\providecommand \@@endlink[0]{}%
\providecommand \url  [0]{\begingroup\@sanitize@url \@url }%
\providecommand \@url [1]{\endgroup\@href {#1}{\urlprefix }}%
\providecommand \urlprefix  [0]{URL }%
\providecommand \Eprint [0]{\href }%
\providecommand \doibase [0]{http://dx.doi.org/}%
\providecommand \selectlanguage [0]{\@gobble}%
\providecommand \bibinfo  [0]{\@secondoftwo}%
\providecommand \bibfield  [0]{\@secondoftwo}%
\providecommand \translation [1]{[#1]}%
\providecommand \BibitemOpen [0]{}%
\providecommand \bibitemStop [0]{}%
\providecommand \bibitemNoStop [0]{.\EOS\space}%
\providecommand \EOS [0]{\spacefactor3000\relax}%
\providecommand \BibitemShut  [1]{\csname bibitem#1\endcsname}%
\let\auto@bib@innerbib\@empty
\bibitem [{\citenamefont {Mueller}\ \emph {et~al.}(2013)\citenamefont
  {Mueller}, \citenamefont {Blackledge},\ and\ \citenamefont
  {Libowitz}}]{mueller2013metal}%
  \BibitemOpen
  \bibfield  {author} {\bibinfo {author} {\bibfnamefont {W.~M.}\ \bibnamefont
  {Mueller}}, \bibinfo {author} {\bibfnamefont {J.~P.}\ \bibnamefont
  {Blackledge}}, \ and\ \bibinfo {author} {\bibfnamefont {G.~G.}\ \bibnamefont
  {Libowitz}},\ }\href@noop {} {\emph {\bibinfo {title} {Metal hydrides}}}\
  (\bibinfo  {publisher} {Elsevier},\ \bibinfo {year} {2013})\BibitemShut
  {NoStop}%
\bibitem [{\citenamefont {Sakintuna}\ \emph {et~al.}(2007)\citenamefont
  {Sakintuna}, \citenamefont {Lamari-Darkrim},\ and\ \citenamefont
  {Hirscher}}]{sakintuna2007metal}%
  \BibitemOpen
  \bibfield  {author} {\bibinfo {author} {\bibfnamefont {B.}~\bibnamefont
  {Sakintuna}}, \bibinfo {author} {\bibfnamefont {F.}~\bibnamefont
  {Lamari-Darkrim}}, \ and\ \bibinfo {author} {\bibfnamefont {M.}~\bibnamefont
  {Hirscher}},\ }\href {\doibase 10.1016/j.ijhydene.2006.11.022} {\bibfield
  {journal} {\bibinfo  {journal} {Int. J. Hydrog. Energy}\ }\textbf {\bibinfo
  {volume} {32}},\ \bibinfo {pages} {1121} (\bibinfo {year}
  {2007})}\BibitemShut {NoStop}%
\bibitem [{\citenamefont {Fukai}(2006)}]{fukai2006metal}%
  \BibitemOpen
  \bibfield  {author} {\bibinfo {author} {\bibfnamefont {Y.}~\bibnamefont
  {Fukai}},\ }\href@noop {} {\emph {\bibinfo {title} {The metal-hydrogen
  system: basic bulk properties}}},\ Vol.~\bibinfo {volume} {21}\ (\bibinfo
  {publisher} {Springer Science \& Business Media},\ \bibinfo {year}
  {2006})\BibitemShut {NoStop}%
\bibitem [{\citenamefont {Satterthwaite}\ and\ \citenamefont
  {Toepke}(1970)}]{satterthwaite1970superconductivity}%
  \BibitemOpen
  \bibfield  {author} {\bibinfo {author} {\bibfnamefont {C.}~\bibnamefont
  {Satterthwaite}}\ and\ \bibinfo {author} {\bibfnamefont {I.}~\bibnamefont
  {Toepke}},\ }\href {\doibase 10.1103/PhysRevLett.25.741} {\bibfield
  {journal} {\bibinfo  {journal} {Phys. Rev. Lett.}\ }\textbf {\bibinfo
  {volume} {25}},\ \bibinfo {pages} {741} (\bibinfo {year} {1970})}\BibitemShut
  {NoStop}%
\bibitem [{\citenamefont
  {Skoskiewicz}(1972)}]{skoskiewicz1972superconductivity}%
  \BibitemOpen
  \bibfield  {author} {\bibinfo {author} {\bibfnamefont {T.}~\bibnamefont
  {Skoskiewicz}},\ }\href
  {https://onlinelibrary.wiley.com/doi/abs/10.1002/pssa.2210110253} {\bibfield
  {journal} {\bibinfo  {journal} {Phys. Status Solidi A}\ }\textbf {\bibinfo
  {volume} {11}} (\bibinfo {year} {1972})}\BibitemShut {NoStop}%
\bibitem [{\citenamefont {Errea}\ \emph {et~al.}(2015)\citenamefont {Errea},
  \citenamefont {Calandra}, \citenamefont {Pickard}, \citenamefont {Nelson},
  \citenamefont {Needs}, \citenamefont {Li}, \citenamefont {Liu}, \citenamefont
  {Zhang}, \citenamefont {Ma},\ and\ \citenamefont {Mauri}}]{errea2015high}%
  \BibitemOpen
  \bibfield  {author} {\bibinfo {author} {\bibfnamefont {I.}~\bibnamefont
  {Errea}}, \bibinfo {author} {\bibfnamefont {M.}~\bibnamefont {Calandra}},
  \bibinfo {author} {\bibfnamefont {C.~J.}\ \bibnamefont {Pickard}}, \bibinfo
  {author} {\bibfnamefont {J.}~\bibnamefont {Nelson}}, \bibinfo {author}
  {\bibfnamefont {R.~J.}\ \bibnamefont {Needs}}, \bibinfo {author}
  {\bibfnamefont {Y.}~\bibnamefont {Li}}, \bibinfo {author} {\bibfnamefont
  {H.}~\bibnamefont {Liu}}, \bibinfo {author} {\bibfnamefont {Y.}~\bibnamefont
  {Zhang}}, \bibinfo {author} {\bibfnamefont {Y.}~\bibnamefont {Ma}}, \ and\
  \bibinfo {author} {\bibfnamefont {F.}~\bibnamefont {Mauri}},\ }\href
  {\doibase 10.1103/PhysRevLett.114.157004} {\bibfield  {journal} {\bibinfo
  {journal} {Phys. Rev. Lett.}\ }\textbf {\bibinfo {volume} {114}},\ \bibinfo
  {pages} {157004} (\bibinfo {year} {2015})}\BibitemShut {NoStop}%
\bibitem [{\citenamefont {Drozdov}\ \emph {et~al.}(2015)\citenamefont
  {Drozdov}, \citenamefont {Eremets}, \citenamefont {Troyan}, \citenamefont
  {Ksenofontov},\ and\ \citenamefont {Shylin}}]{drozdov2015conventional}%
  \BibitemOpen
  \bibfield  {author} {\bibinfo {author} {\bibfnamefont {A.}~\bibnamefont
  {Drozdov}}, \bibinfo {author} {\bibfnamefont {M.}~\bibnamefont {Eremets}},
  \bibinfo {author} {\bibfnamefont {I.}~\bibnamefont {Troyan}}, \bibinfo
  {author} {\bibfnamefont {V.}~\bibnamefont {Ksenofontov}}, \ and\ \bibinfo
  {author} {\bibfnamefont {S.}~\bibnamefont {Shylin}},\ }\href {\doibase
  10.1038/nature14964} {\bibfield  {journal} {\bibinfo  {journal} {Nature}\
  }\textbf {\bibinfo {volume} {525}},\ \bibinfo {pages} {73} (\bibinfo {year}
  {2015})}\BibitemShut {NoStop}%
\bibitem [{\citenamefont {Peng}\ \emph {et~al.}(2017)\citenamefont {Peng},
  \citenamefont {Sun}, \citenamefont {Pickard}, \citenamefont {Needs},
  \citenamefont {Wu},\ and\ \citenamefont {Ma}}]{peng2017hydrogen}%
  \BibitemOpen
  \bibfield  {author} {\bibinfo {author} {\bibfnamefont {F.}~\bibnamefont
  {Peng}}, \bibinfo {author} {\bibfnamefont {Y.}~\bibnamefont {Sun}}, \bibinfo
  {author} {\bibfnamefont {C.~J.}\ \bibnamefont {Pickard}}, \bibinfo {author}
  {\bibfnamefont {R.~J.}\ \bibnamefont {Needs}}, \bibinfo {author}
  {\bibfnamefont {Q.}~\bibnamefont {Wu}}, \ and\ \bibinfo {author}
  {\bibfnamefont {Y.}~\bibnamefont {Ma}},\ }\href {\doibase
  10.1103/PhysRevLett.119.107001} {\bibfield  {journal} {\bibinfo  {journal}
  {Phys. Rev. Lett.}\ }\textbf {\bibinfo {volume} {119}},\ \bibinfo {pages}
  {107001} (\bibinfo {year} {2017})}\BibitemShut {NoStop}%
\bibitem [{\citenamefont {Drozdov}\ \emph {et~al.}()\citenamefont {Drozdov},
  \citenamefont {Kong}, \citenamefont {Minkov}, \citenamefont {Besedin},
  \citenamefont {Kuzovnikov}, \citenamefont {Mozaffari}, \citenamefont
  {Balicas}, \citenamefont {Balakirev}, \citenamefont {Graf}, \citenamefont
  {Prakapenka} \emph {et~al.}}]{drozdov2018superconductivity}%
  \BibitemOpen
  \bibfield  {author} {\bibinfo {author} {\bibfnamefont {A.}~\bibnamefont
  {Drozdov}}, \bibinfo {author} {\bibfnamefont {P.}~\bibnamefont {Kong}},
  \bibinfo {author} {\bibfnamefont {V.}~\bibnamefont {Minkov}}, \bibinfo
  {author} {\bibfnamefont {S.}~\bibnamefont {Besedin}}, \bibinfo {author}
  {\bibfnamefont {M.}~\bibnamefont {Kuzovnikov}}, \bibinfo {author}
  {\bibfnamefont {S.}~\bibnamefont {Mozaffari}}, \bibinfo {author}
  {\bibfnamefont {L.}~\bibnamefont {Balicas}}, \bibinfo {author} {\bibfnamefont
  {F.}~\bibnamefont {Balakirev}}, \bibinfo {author} {\bibfnamefont
  {D.}~\bibnamefont {Graf}}, \bibinfo {author} {\bibfnamefont {V.}~\bibnamefont
  {Prakapenka}},  \emph {et~al.},\ }\href {\doibase
  10.1038/s41586-019-1201-8} {\bibfield  {journal} {\bibinfo  {journal} {Nature}\
  }\textbf {\bibinfo {volume} {569}},\ \bibinfo {pages} {528} (\bibinfo {year}
  {2019})}\BibitemShut {NoStop}%
\bibitem [{\citenamefont {Geerken}\ and\ \citenamefont
  {Griessen}(1983)}]{geerken1983concentration}%
  \BibitemOpen
  \bibfield  {author} {\bibinfo {author} {\bibfnamefont {B.}~\bibnamefont
  {Geerken}}\ and\ \bibinfo {author} {\bibfnamefont {R.}~\bibnamefont
  {Griessen}},\ }\href {\doibase 10.1088/0305-4608/13/5/010} {\bibfield
  {journal} {\bibinfo  {journal} {J. Phys. F: Met. Phys.}\ }\textbf {\bibinfo
  {volume} {13}},\ \bibinfo {pages} {963} (\bibinfo {year} {1983})}\BibitemShut
  {NoStop}%
\bibitem [{\citenamefont {Sakamoto}\ \emph {et~al.}(1996)\citenamefont
  {Sakamoto}, \citenamefont {Takai}, \citenamefont {Takashima},\ and\
  \citenamefont {Imada}}]{sakamoto1996electrical}%
  \BibitemOpen
  \bibfield  {author} {\bibinfo {author} {\bibfnamefont {Y.}~\bibnamefont
  {Sakamoto}}, \bibinfo {author} {\bibfnamefont {K.}~\bibnamefont {Takai}},
  \bibinfo {author} {\bibfnamefont {I.}~\bibnamefont {Takashima}}, \ and\
  \bibinfo {author} {\bibfnamefont {M.}~\bibnamefont {Imada}},\ }\href
  {\doibase 10.1088/0953-8984/8/19/015} {\bibfield  {journal} {\bibinfo
  {journal} {J. Phys.: Condens. Matter}\ }\textbf {\bibinfo {volume} {8}},\
  \bibinfo {pages} {3399} (\bibinfo {year} {1996})}\BibitemShut {NoStop}%
\bibitem [{\citenamefont {Bickel}\ and\ \citenamefont
  {Berlincourt}(1970)}]{bickel1970electrical}%
  \BibitemOpen
  \bibfield  {author} {\bibinfo {author} {\bibfnamefont {P.}~\bibnamefont
  {Bickel}}\ and\ \bibinfo {author} {\bibfnamefont {T.}~\bibnamefont
  {Berlincourt}},\ }\href {\doibase 10.1103/PhysRevB.2.4807} {\bibfield
  {journal} {\bibinfo  {journal} {Phys. Rev. B}\ }\textbf {\bibinfo {volume}
  {2}},\ \bibinfo {pages} {4807} (\bibinfo {year} {1970})}\BibitemShut
  {NoStop}%
\bibitem [{\citenamefont {Wang}\ and\ \citenamefont
  {Gong}(2012{\natexlab{a}})}]{wang2012first}%
  \BibitemOpen
  \bibfield  {author} {\bibinfo {author} {\bibfnamefont {F.}~\bibnamefont
  {Wang}}\ and\ \bibinfo {author} {\bibfnamefont {H.}~\bibnamefont {Gong}},\
  }\href {\doibase 10.1016/j.ijhydene.2012.06.037} {\bibfield  {journal}
  {\bibinfo  {journal} {Int. J. Hydrog. Energy}\ }\textbf {\bibinfo {volume}
  {37}},\ \bibinfo {pages} {12393} (\bibinfo {year}
  {2012}{\natexlab{a}})}\BibitemShut {NoStop}%
\bibitem [{\citenamefont {Zieli{\'n}ski}\ and\ \citenamefont
  {Sobieszczyk}(2011)}]{zielinski2011hydrogen}%
  \BibitemOpen
  \bibfield  {author} {\bibinfo {author} {\bibfnamefont {A.}~\bibnamefont
  {Zieli{\'n}ski}}\ and\ \bibinfo {author} {\bibfnamefont {S.}~\bibnamefont
  {Sobieszczyk}},\ }\href {\doibase 10.1016/j.ijhydene.2011.04.002} {\bibfield
  {journal} {\bibinfo  {journal} {Int. J. Hydrog. Energy}\ }\textbf {\bibinfo
  {volume} {36}},\ \bibinfo {pages} {8619} (\bibinfo {year}
  {2011})}\BibitemShut {NoStop}%
\bibitem [{\citenamefont {Zuzek}\ \emph {et~al.}(1990)\citenamefont {Zuzek},
  \citenamefont {Abriata}, \citenamefont {San-Martin},\ and\ \citenamefont
  {Manchester}}]{zuzek1990h}%
  \BibitemOpen
  \bibfield  {author} {\bibinfo {author} {\bibfnamefont {E.}~\bibnamefont
  {Zuzek}}, \bibinfo {author} {\bibfnamefont {J.}~\bibnamefont {Abriata}},
  \bibinfo {author} {\bibfnamefont {A.}~\bibnamefont {San-Martin}}, \ and\
  \bibinfo {author} {\bibfnamefont {F.}~\bibnamefont {Manchester}},\ }\href
  {https://link.springer.com/content/pdf/10.1007/BF02843318.pdf} {\bibfield
  {journal} {\bibinfo  {journal} {Bulletin of Alloy Phase Diagrams}\ }\textbf
  {\bibinfo {volume} {11}},\ \bibinfo {pages} {385} (\bibinfo {year}
  {1990})}\BibitemShut {NoStop}%
\bibitem [{\citenamefont {Grib}\ \emph {et~al.}(2010)\citenamefont {Grib},
  \citenamefont {Khadzhay}, \citenamefont {Merisov}, \citenamefont
  {Vinogradov},\ and\ \citenamefont {Tikhonovsky}}]{grib2010kinetics}%
  \BibitemOpen
  \bibfield  {author} {\bibinfo {author} {\bibfnamefont {A.}~\bibnamefont
  {Grib}}, \bibinfo {author} {\bibfnamefont {G.}~\bibnamefont {Khadzhay}},
  \bibinfo {author} {\bibfnamefont {B.}~\bibnamefont {Merisov}}, \bibinfo
  {author} {\bibfnamefont {D.}~\bibnamefont {Vinogradov}}, \ and\ \bibinfo
  {author} {\bibfnamefont {M.}~\bibnamefont {Tikhonovsky}},\ }\href {\doibase
  10.1016/j.ijhydene.2010.02.139} {\bibfield  {journal} {\bibinfo  {journal}
  {Int. J. Hydrog. Energy}\ }\textbf {\bibinfo {volume} {35}},\ \bibinfo
  {pages} {5442} (\bibinfo {year} {2010})}\BibitemShut {NoStop}%
\bibitem [{\citenamefont {Tsuchiya}\ \emph
  {et~al.}(2002{\natexlab{a}})\citenamefont {Tsuchiya}, \citenamefont
  {Teshigawara}, \citenamefont {Konashi},\ and\ \citenamefont
  {Yamawaki}}]{tsuchiya2002thermal}%
  \BibitemOpen
  \bibfield  {author} {\bibinfo {author} {\bibfnamefont {B.}~\bibnamefont
  {Tsuchiya}}, \bibinfo {author} {\bibfnamefont {M.}~\bibnamefont
  {Teshigawara}}, \bibinfo {author} {\bibfnamefont {K.}~\bibnamefont
  {Konashi}}, \ and\ \bibinfo {author} {\bibfnamefont {M.}~\bibnamefont
  {Yamawaki}},\ }\href {\doibase 10.1016/S0925-8388(01)01434-7} {\bibfield
  {journal} {\bibinfo  {journal} {J. Alloy. Compd.}\ }\textbf {\bibinfo
  {volume} {330}},\ \bibinfo {pages} {357} (\bibinfo {year}
  {2002}{\natexlab{a}})}\BibitemShut {NoStop}%
\bibitem [{\citenamefont {Uno}\ \emph {et~al.}(2004)\citenamefont {Uno},
  \citenamefont {Yamada}, \citenamefont {Maruyama}, \citenamefont {Muta},\ and\
  \citenamefont {Yamanaka}}]{uno2004thermophysical}%
  \BibitemOpen
  \bibfield  {author} {\bibinfo {author} {\bibfnamefont {M.}~\bibnamefont
  {Uno}}, \bibinfo {author} {\bibfnamefont {K.}~\bibnamefont {Yamada}},
  \bibinfo {author} {\bibfnamefont {T.}~\bibnamefont {Maruyama}}, \bibinfo
  {author} {\bibfnamefont {H.}~\bibnamefont {Muta}}, \ and\ \bibinfo {author}
  {\bibfnamefont {S.}~\bibnamefont {Yamanaka}},\ }\href {\doibase
  10.1016/j.jallcom.2003.07.006} {\bibfield  {journal} {\bibinfo  {journal} {J.
  Alloy. Compd.}\ }\textbf {\bibinfo {volume} {366}},\ \bibinfo {pages} {101}
  (\bibinfo {year} {2004})}\BibitemShut {NoStop}%
\bibitem [{\citenamefont {Tsuchiya}\ \emph
  {et~al.}(2002{\natexlab{b}})\citenamefont {Tsuchiya}, \citenamefont
  {Teshigawara}, \citenamefont {Konashi}, \citenamefont {Nagata}, \citenamefont
  {Shikama},\ and\ \citenamefont {Yamawaki}}]{tsuchiya2002isotope}%
  \BibitemOpen
  \bibfield  {author} {\bibinfo {author} {\bibfnamefont {B.}~\bibnamefont
  {Tsuchiya}}, \bibinfo {author} {\bibfnamefont {M.}~\bibnamefont
  {Teshigawara}}, \bibinfo {author} {\bibfnamefont {K.}~\bibnamefont
  {Konashi}}, \bibinfo {author} {\bibfnamefont {S.}~\bibnamefont {Nagata}},
  \bibinfo {author} {\bibfnamefont {T.}~\bibnamefont {Shikama}}, \ and\
  \bibinfo {author} {\bibfnamefont {M.}~\bibnamefont {Yamawaki}},\ }\href
  {\doibase 10.1080/18811248.2002.9715213} {\bibfield  {journal} {\bibinfo
  {journal} {J. Nucl. Sci. Technol.}\ }\textbf {\bibinfo {volume} {39}},\
  \bibinfo {pages} {402} (\bibinfo {year} {2002}{\natexlab{b}})}\BibitemShut
  {NoStop}%
\bibitem [{\citenamefont {Khoda-Bakhsh}\ and\ \citenamefont
  {Ross}(1982)}]{khoda1982determination}%
  \BibitemOpen
  \bibfield  {author} {\bibinfo {author} {\bibfnamefont {R.}~\bibnamefont
  {Khoda-Bakhsh}}\ and\ \bibinfo {author} {\bibfnamefont {D.}~\bibnamefont
  {Ross}},\ }\href {\doibase 10.1088/0305-4608/12/1/003} {\bibfield  {journal}
  {\bibinfo  {journal} {J. Phys. F: Met. Phys.}\ }\textbf {\bibinfo {volume}
  {12}},\ \bibinfo {pages} {15} (\bibinfo {year} {1982})}\BibitemShut {NoStop}%
\bibitem [{\citenamefont {Narang}\ \emph {et~al.}(1977)\citenamefont {Narang},
  \citenamefont {Paul},\ and\ \citenamefont {Taylor}}]{narang1977location}%
  \BibitemOpen
  \bibfield  {author} {\bibinfo {author} {\bibfnamefont {P.}~\bibnamefont
  {Narang}}, \bibinfo {author} {\bibfnamefont {G.}~\bibnamefont {Paul}}, \ and\
  \bibinfo {author} {\bibfnamefont {K.}~\bibnamefont {Taylor}},\ }\href
  {\doibase 10.1016/0022-5088(77)90225-9} {\bibfield  {journal} {\bibinfo
  {journal} {Journal of the Less Common Metals}\ }\textbf {\bibinfo {volume}
  {56}},\ \bibinfo {pages} {125} (\bibinfo {year} {1977})}\BibitemShut
  {NoStop}%
\bibitem [{\citenamefont {Lopatina}\ \emph {et~al.}(2015)\citenamefont
  {Lopatina}, \citenamefont {Svyatkin}, \citenamefont {Koroteev},\ and\
  \citenamefont {Chernov}}]{lopatina2015electronic}%
  \BibitemOpen
  \bibfield  {author} {\bibinfo {author} {\bibfnamefont {O.~V.}\ \bibnamefont
  {Lopatina}}, \bibinfo {author} {\bibfnamefont {L.~A.}\ \bibnamefont
  {Svyatkin}}, \bibinfo {author} {\bibfnamefont {Y.~M.}\ \bibnamefont
  {Koroteev}}, \ and\ \bibinfo {author} {\bibfnamefont {I.~P.}\ \bibnamefont
  {Chernov}},\ }\href {\doibase 10.1134/S106378341509022X} {\bibfield
  {journal} {\bibinfo  {journal} {Phys. Solid State}\ }\textbf {\bibinfo
  {volume} {57}},\ \bibinfo {pages} {1719} (\bibinfo {year}
  {2015})}\BibitemShut {NoStop}%
\bibitem [{\citenamefont {Svyatkin}\ \emph {et~al.}(2018)\citenamefont
  {Svyatkin}, \citenamefont {Koroteev},\ and\ \citenamefont
  {Chernov}}]{svyatkin2018mutual}%
  \BibitemOpen
  \bibfield  {author} {\bibinfo {author} {\bibfnamefont {L.}~\bibnamefont
  {Svyatkin}}, \bibinfo {author} {\bibfnamefont {Y.~M.}\ \bibnamefont
  {Koroteev}}, \ and\ \bibinfo {author} {\bibfnamefont {I.}~\bibnamefont
  {Chernov}},\ }\href {\doibase 10.1134/S1063783418010262} {\bibfield
  {journal} {\bibinfo  {journal} {Phys. Solid State}\ }\textbf {\bibinfo
  {volume} {60}},\ \bibinfo {pages} {10} (\bibinfo {year} {2018})}\BibitemShut
  {NoStop}%
\bibitem [{\citenamefont {Ackland}(1998)}]{ackland1998embrittlement}%
  \BibitemOpen
  \bibfield  {author} {\bibinfo {author} {\bibfnamefont {G.}~\bibnamefont
  {Ackland}},\ }\href {\doibase 10.1103/PhysRevLett.81.3300} {\bibfield
  {journal} {\bibinfo  {journal} {Phys. Rev. Lett.}\ }\textbf {\bibinfo
  {volume} {80}},\ \bibinfo {pages} {2233} (\bibinfo {year}
  {1998})}\BibitemShut {NoStop}%
\bibitem [{\citenamefont {Nied{\'z}wied{\'z}}\ \emph
  {et~al.}(1993)\citenamefont {Nied{\'z}wied{\'z}}, \citenamefont {Nowak} \emph
  {et~al.}}]{niedzwiedz199391zr}%
  \BibitemOpen
  \bibfield  {author} {\bibinfo {author} {\bibfnamefont {K.}~\bibnamefont
  {Nied{\'z}wied{\'z}}}, \bibinfo {author} {\bibfnamefont {B.}~\bibnamefont
  {Nowak}},  \emph {et~al.},\ }\href {\doibase 10.1016/0925-8388(93)90643-2}
  {\bibfield  {journal} {\bibinfo  {journal} {J. Alloy. Compd.}\ }\textbf
  {\bibinfo {volume} {194}},\ \bibinfo {pages} {47} (\bibinfo {year}
  {1993})}\BibitemShut {NoStop}%
\bibitem [{\citenamefont {Cantrell}\ \emph {et~al.}(1984)\citenamefont
  {Cantrell}, \citenamefont {Bowman},\ and\ \citenamefont
  {Sullenger}}]{cantrell1984x}%
  \BibitemOpen
  \bibfield  {author} {\bibinfo {author} {\bibfnamefont {J.~S.}\ \bibnamefont
  {Cantrell}}, \bibinfo {author} {\bibfnamefont {R.~J.}\ \bibnamefont
  {Bowman}}, \ and\ \bibinfo {author} {\bibfnamefont {D.}~\bibnamefont
  {Sullenger}},\ }\href {\doibase 10.1021/j150649a018} {\bibfield  {journal}
  {\bibinfo  {journal} {J. Phys. Chem.}\ }\textbf {\bibinfo {volume} {88}},\
  \bibinfo {pages} {918} (\bibinfo {year} {1984})}\BibitemShut {NoStop}%
\bibitem [{\citenamefont {Bowman~Jr}\ \emph {et~al.}(1983)\citenamefont
  {Bowman~Jr}, \citenamefont {Venturini}, \citenamefont {Craft}, \citenamefont
  {Attalla},\ and\ \citenamefont {Sullenger}}]{bowman1983electronic}%
  \BibitemOpen
  \bibfield  {author} {\bibinfo {author} {\bibfnamefont {R.}~\bibnamefont
  {Bowman~Jr}}, \bibinfo {author} {\bibfnamefont {E.}~\bibnamefont
  {Venturini}}, \bibinfo {author} {\bibfnamefont {B.}~\bibnamefont {Craft}},
  \bibinfo {author} {\bibfnamefont {A.}~\bibnamefont {Attalla}}, \ and\
  \bibinfo {author} {\bibfnamefont {D.}~\bibnamefont {Sullenger}},\ }\href
  {\doibase 10.1103/PhysRevB.27.1474} {\bibfield  {journal} {\bibinfo
  {journal} {Phys. Rev. B}\ }\textbf {\bibinfo {volume} {27}},\ \bibinfo
  {pages} {1474} (\bibinfo {year} {1983})}\BibitemShut {NoStop}%
\bibitem [{\citenamefont {Yakel}(1958)}]{yakel1958thermocrystallography}%
  \BibitemOpen
  \bibfield  {author} {\bibinfo {author} {\bibfnamefont {H.}~\bibnamefont
  {Yakel}},\ }\href {\doibase 10.1107/S0365110X58000098} {\bibfield  {journal}
  {\bibinfo  {journal} {Acta Crystallogr.}\ }\textbf {\bibinfo {volume} {11}},\
  \bibinfo {pages} {46} (\bibinfo {year} {1958})}\BibitemShut {NoStop}%
\bibitem [{\citenamefont {Kul’kova}\ \emph {et~al.}(1999)\citenamefont
  {Kul’kova}, \citenamefont {Muryzhnikova},\ and\ \citenamefont
  {Naumov}}]{kul1999electronic}%
  \BibitemOpen
  \bibfield  {author} {\bibinfo {author} {\bibfnamefont {S.}~\bibnamefont
  {Kul’kova}}, \bibinfo {author} {\bibfnamefont {O.}~\bibnamefont
  {Muryzhnikova}}, \ and\ \bibinfo {author} {\bibfnamefont {I.}~\bibnamefont
  {Naumov}},\ }\href {\doibase 10.1134/1.1131093} {\bibfield  {journal}
  {\bibinfo  {journal} {Phys. Solid State}\ }\textbf {\bibinfo {volume} {41}},\
  \bibinfo {pages} {1763} (\bibinfo {year} {1999})}\BibitemShut {NoStop}%
\bibitem [{\citenamefont {Miwa}\ and\ \citenamefont
  {Fukumoto}(2002)}]{miwa2002first}%
  \BibitemOpen
  \bibfield  {author} {\bibinfo {author} {\bibfnamefont {K.}~\bibnamefont
  {Miwa}}\ and\ \bibinfo {author} {\bibfnamefont {A.}~\bibnamefont
  {Fukumoto}},\ }\href {\doibase 10.1103/PhysRevB.65.155114} {\bibfield
  {journal} {\bibinfo  {journal} {Phys. Rev. B}\ }\textbf {\bibinfo {volume}
  {65}},\ \bibinfo {pages} {155114} (\bibinfo {year} {2002})}\BibitemShut
  {NoStop}%
\bibitem [{\citenamefont {Quijano}\ \emph {et~al.}(2009)\citenamefont
  {Quijano}, \citenamefont {de~Coss},\ and\ \citenamefont
  {Singh}}]{quijano2009electronic}%
  \BibitemOpen
  \bibfield  {author} {\bibinfo {author} {\bibfnamefont {R.}~\bibnamefont
  {Quijano}}, \bibinfo {author} {\bibfnamefont {R.}~\bibnamefont {de~Coss}}, \
  and\ \bibinfo {author} {\bibfnamefont {D.~J.}\ \bibnamefont {Singh}},\ }\href
  {\doibase 10.1103/PhysRevB.80.184103} {\bibfield  {journal} {\bibinfo
  {journal} {Phys. Rev. B}\ }\textbf {\bibinfo {volume} {80}},\ \bibinfo
  {pages} {184103} (\bibinfo {year} {2009})}\BibitemShut {NoStop}%
\bibitem [{\citenamefont {Gupta}(1982)}]{gupta1982electronic}%
  \BibitemOpen
  \bibfield  {author} {\bibinfo {author} {\bibfnamefont {M.}~\bibnamefont
  {Gupta}},\ }\href {\doibase 10.1103/PhysRevB.25.1027} {\bibfield  {journal}
  {\bibinfo  {journal} {Phys. Rev. B}\ }\textbf {\bibinfo {volume} {25}},\
  \bibinfo {pages} {1027} (\bibinfo {year} {1982})}\BibitemShut {NoStop}%
\bibitem [{\citenamefont {Wolf}\ and\ \citenamefont
  {Herzig}(2000)}]{wolf2000first}%
  \BibitemOpen
  \bibfield  {author} {\bibinfo {author} {\bibfnamefont {W.}~\bibnamefont
  {Wolf}}\ and\ \bibinfo {author} {\bibfnamefont {P.}~\bibnamefont {Herzig}},\
  }\href {\doibase 10.1088/0953-8984/12/21/301} {\bibfield  {journal} {\bibinfo
   {journal} {J. Phys.: Condens. Matter}\ }\textbf {\bibinfo {volume} {12}},\
  \bibinfo {pages} {4535} (\bibinfo {year} {2000})}\BibitemShut {NoStop}%
\bibitem [{\citenamefont {Zhang}\ \emph {et~al.}(2011)\citenamefont {Zhang},
  \citenamefont {Wang}, \citenamefont {He},\ and\ \citenamefont
  {Zhang}}]{zhang2011first}%
  \BibitemOpen
  \bibfield  {author} {\bibinfo {author} {\bibfnamefont {P.}~\bibnamefont
  {Zhang}}, \bibinfo {author} {\bibfnamefont {B.-T.}\ \bibnamefont {Wang}},
  \bibinfo {author} {\bibfnamefont {C.-H.}\ \bibnamefont {He}}, \ and\ \bibinfo
  {author} {\bibfnamefont {P.}~\bibnamefont {Zhang}},\ }\href {\doibase
  10.1016/j.commatsci.2011.06.016} {\bibfield  {journal} {\bibinfo  {journal}
  {Comp. Mater. Sci.}\ }\textbf {\bibinfo {volume} {50}},\ \bibinfo {pages}
  {3297} (\bibinfo {year} {2011})}\BibitemShut {NoStop}%
\bibitem [{\citenamefont {Wang}\ and\ \citenamefont
  {Gong}(2012{\natexlab{b}})}]{wang2012mechanical}%
  \BibitemOpen
  \bibfield  {author} {\bibinfo {author} {\bibfnamefont {F.}~\bibnamefont
  {Wang}}\ and\ \bibinfo {author} {\bibfnamefont {H.}~\bibnamefont {Gong}},\
  }\href {\doibase 10.1016/j.ijhydene.2012.03.078} {\bibfield  {journal}
  {\bibinfo  {journal} {Int. J. Hydrog. Energy}\ }\textbf {\bibinfo {volume}
  {37}},\ \bibinfo {pages} {9688} (\bibinfo {year}
  {2012}{\natexlab{b}})}\BibitemShut {NoStop}%
\bibitem [{\citenamefont {Hamann}(2013)}]{hamann2013optimized}%
  \BibitemOpen
  \bibfield  {author} {\bibinfo {author} {\bibfnamefont {D.}~\bibnamefont
  {Hamann}},\ }\href {\doibase 10.1103/PhysRevB.88.085117} {\bibfield
  {journal} {\bibinfo  {journal} {Phys. Rev. B}\ }\textbf {\bibinfo {volume}
  {88}},\ \bibinfo {pages} {085117} (\bibinfo {year} {2013})}\BibitemShut
  {NoStop}%
\bibitem [{\citenamefont {Gonze}\ \emph {et~al.}(2016)\citenamefont {Gonze},
  \citenamefont {Jollet}, \citenamefont {Araujo}, \citenamefont {Adams},
  \citenamefont {Amadon}, \citenamefont {Applencourt}, \citenamefont {Audouze},
  \citenamefont {Beuken}, \citenamefont {Bieder}, \citenamefont {Bokhanchuk}
  \emph {et~al.}}]{gonze2016recent}%
  \BibitemOpen
  \bibfield  {author} {\bibinfo {author} {\bibfnamefont {X.}~\bibnamefont
  {Gonze}}, \bibinfo {author} {\bibfnamefont {F.}~\bibnamefont {Jollet}},
  \bibinfo {author} {\bibfnamefont {F.~A.}\ \bibnamefont {Araujo}}, \bibinfo
  {author} {\bibfnamefont {D.}~\bibnamefont {Adams}}, \bibinfo {author}
  {\bibfnamefont {B.}~\bibnamefont {Amadon}}, \bibinfo {author} {\bibfnamefont
  {T.}~\bibnamefont {Applencourt}}, \bibinfo {author} {\bibfnamefont
  {C.}~\bibnamefont {Audouze}}, \bibinfo {author} {\bibfnamefont {J.-M.}\
  \bibnamefont {Beuken}}, \bibinfo {author} {\bibfnamefont {J.}~\bibnamefont
  {Bieder}}, \bibinfo {author} {\bibfnamefont {A.}~\bibnamefont {Bokhanchuk}},
  \emph {et~al.},\ }\href {\doibase 10.1016/j.cpc.2016.04.003} {\bibfield
  {journal} {\bibinfo  {journal} {Comput. Phys. Commun.}\ }\textbf {\bibinfo
  {volume} {205}},\ \bibinfo {pages} {106} (\bibinfo {year}
  {2016})}\BibitemShut {NoStop}%
\bibitem [{\citenamefont {Perdew}\ \emph {et~al.}(1996)\citenamefont {Perdew},
  \citenamefont {Burke},\ and\ \citenamefont
  {Ernzerhof}}]{perdew1996generalized}%
  \BibitemOpen
  \bibfield  {author} {\bibinfo {author} {\bibfnamefont {J.~P.}\ \bibnamefont
  {Perdew}}, \bibinfo {author} {\bibfnamefont {K.}~\bibnamefont {Burke}}, \
  and\ \bibinfo {author} {\bibfnamefont {M.}~\bibnamefont {Ernzerhof}},\ }\href
  {\doibase 10.1103/PhysRevLett.77.3865} {\bibfield  {journal} {\bibinfo
  {journal} {Phys. Rev. Lett.}\ }\textbf {\bibinfo {volume} {77}},\ \bibinfo
  {pages} {3865} (\bibinfo {year} {1996})}\BibitemShut {NoStop}%
\bibitem [{\citenamefont {Verstraete}\ and\ \citenamefont
  {Gonze}(2002)}]{verstraete2002}%
  \BibitemOpen
  \bibfield  {author} {\bibinfo {author} {\bibfnamefont {M.}~\bibnamefont
  {Verstraete}}\ and\ \bibinfo {author} {\bibfnamefont {X.}~\bibnamefont
  {Gonze}},\ }\href {\doibase 10.1103/PhysRevB.65.035111} {\bibfield  {journal}
  {\bibinfo  {journal} {Phys. Rev. B}\ }\textbf {\bibinfo {volume} {65}},\
  \bibinfo {pages} {1} (\bibinfo {year} {2002})}\BibitemShut {NoStop}%
\bibitem [{\citenamefont {Eitan}\ \emph {et~al.}(2011)\citenamefont {Eitan},
  \citenamefont {Mundt},\ and\ \citenamefont {Tannor}}]{eitan2011optimal}%
  \BibitemOpen
  \bibfield  {author} {\bibinfo {author} {\bibfnamefont {R.}~\bibnamefont
  {Eitan}}, \bibinfo {author} {\bibfnamefont {M.}~\bibnamefont {Mundt}}, \ and\
  \bibinfo {author} {\bibfnamefont {D.~J.}\ \bibnamefont {Tannor}},\ }\href
  {\doibase 10.1103/PhysRevA.83.053426} {\bibfield  {journal} {\bibinfo
  {journal} {Phys. Rev. A}\ }\textbf {\bibinfo {volume} {83}},\ \bibinfo
  {pages} {053426} (\bibinfo {year} {2011})}\BibitemShut {NoStop}%
\bibitem [{\citenamefont {Giustino}(2017)}]{giustino2017electron}%
  \BibitemOpen
  \bibfield  {author} {\bibinfo {author} {\bibfnamefont {F.}~\bibnamefont
  {Giustino}},\ }\href {\doibase 10.1103/RevModPhys.89.015003} {\bibfield
  {journal} {\bibinfo  {journal} {Rev. Mod. Phys.}\ }\textbf {\bibinfo {volume}
  {89}},\ \bibinfo {pages} {015003} (\bibinfo {year} {2017})}\BibitemShut
  {NoStop}%
\bibitem [{\citenamefont {Allen}(1972)}]{allen1972neutron}%
  \BibitemOpen
  \bibfield  {author} {\bibinfo {author} {\bibfnamefont {P.~B.}\ \bibnamefont
  {Allen}},\ }\href {\doibase 10.1103/PhysRevB.6.2577} {\bibfield  {journal}
  {\bibinfo  {journal} {Phys. Rev. B}\ }\textbf {\bibinfo {volume} {6}},\
  \bibinfo {pages} {2577} (\bibinfo {year} {1972})}\BibitemShut {NoStop}%
\bibitem [{\citenamefont {Zaharioudakis}(2004)}]{zaharioudakis2004tetrahedron}%
  \BibitemOpen
  \bibfield  {author} {\bibinfo {author} {\bibfnamefont {D.}~\bibnamefont
  {Zaharioudakis}},\ }\href {\doibase 10.1016/S0010-4655(03)00489-2} {\bibfield
   {journal} {\bibinfo  {journal} {Comput. Phys. Commun.}\ }\textbf {\bibinfo
  {volume} {157}},\ \bibinfo {pages} {17} (\bibinfo {year} {2004})}\BibitemShut
  {NoStop}%
\bibitem [{\citenamefont {Allen}(1971)}]{allen1971electron}%
  \BibitemOpen
  \bibfield  {author} {\bibinfo {author} {\bibfnamefont {P.}~\bibnamefont
  {Allen}},\ }\href {\doibase 10.1103/PhysRevB.3.305} {\bibfield  {journal}
  {\bibinfo  {journal} {Phys. Rev. B}\ }\textbf {\bibinfo {volume} {3}},\
  \bibinfo {pages} {305} (\bibinfo {year} {1971})}\BibitemShut {NoStop}%
\bibitem [{\citenamefont {Savrasov}\ and\ \citenamefont
  {Savrasov}(1996)}]{savrasov1996electron}%
  \BibitemOpen
  \bibfield  {author} {\bibinfo {author} {\bibfnamefont {S.~Y.}\ \bibnamefont
  {Savrasov}}\ and\ \bibinfo {author} {\bibfnamefont {D.~Y.}\ \bibnamefont
  {Savrasov}},\ }\href {\doibase 10.1103/PhysRevB.54.16487} {\bibfield
  {journal} {\bibinfo  {journal} {Phys. Rev. B}\ }\textbf {\bibinfo {volume}
  {54}},\ \bibinfo {pages} {16487} (\bibinfo {year} {1996})}\BibitemShut
  {NoStop}%
\bibitem [{\citenamefont {Desai}\ \emph {et~al.}(1984)\citenamefont {Desai},
  \citenamefont {James},\ and\ \citenamefont {Ho}}]{desai1984electrical}%
  \BibitemOpen
  \bibfield  {author} {\bibinfo {author} {\bibfnamefont {P.~D.}\ \bibnamefont
  {Desai}}, \bibinfo {author} {\bibfnamefont {H.~M.}\ \bibnamefont {James}}, \
  and\ \bibinfo {author} {\bibfnamefont {C.~Y.}\ \bibnamefont {Ho}},\ }\href
  {\doibase 10.1063/1.555724} {\bibfield  {journal} {\bibinfo  {journal} {J.
  Phys. Chem. Ref. Data}\ }\textbf {\bibinfo {volume} {13}},\ \bibinfo {pages}
  {1097} (\bibinfo {year} {1984})}\BibitemShut {NoStop}%
\bibitem [{\citenamefont {Mott}(1964)}]{mott1964electrons}%
  \BibitemOpen
  \bibfield  {author} {\bibinfo {author} {\bibfnamefont {N.~F.}\ \bibnamefont
  {Mott}},\ }\href {\doibase 10.1080/00018736400101041} {\bibfield  {journal}
  {\bibinfo  {journal} {Advances in Physics}\ }\textbf {\bibinfo {volume}
  {13}},\ \bibinfo {pages} {325} (\bibinfo {year} {1964})}\BibitemShut
  {NoStop}%
\bibitem [{\citenamefont {Crespi}\ and\ \citenamefont
  {Cohen}(1992)}]{crespi1992possible}%
  \BibitemOpen
  \bibfield  {author} {\bibinfo {author} {\bibfnamefont {V.~H.}\ \bibnamefont
  {Crespi}}\ and\ \bibinfo {author} {\bibfnamefont {M.~L.}\ \bibnamefont
  {Cohen}},\ }\href {\doibase 10.1016/0038-1098(92)90386-N} {\bibfield
  {journal} {\bibinfo  {journal} {Solid State Commun.}\ }\textbf {\bibinfo
  {volume} {81}},\ \bibinfo {pages} {187} (\bibinfo {year} {1992})}\BibitemShut
  {NoStop}%
\bibitem [{\citenamefont {Giannozzi}\ \emph {et~al.}(1991)\citenamefont
  {Giannozzi}, \citenamefont {De~Gironcoli}, \citenamefont {Pavone},\ and\
  \citenamefont {Baroni}}]{giannozzi1991ab}%
  \BibitemOpen
  \bibfield  {author} {\bibinfo {author} {\bibfnamefont {P.}~\bibnamefont
  {Giannozzi}}, \bibinfo {author} {\bibfnamefont {S.}~\bibnamefont
  {De~Gironcoli}}, \bibinfo {author} {\bibfnamefont {P.}~\bibnamefont
  {Pavone}}, \ and\ \bibinfo {author} {\bibfnamefont {S.}~\bibnamefont
  {Baroni}},\ }\href {\doibase 10.1103/PhysRevB.43.7231} {\bibfield  {journal}
  {\bibinfo  {journal} {Phys. Rev. B}\ }\textbf {\bibinfo {volume} {43}},\
  \bibinfo {pages} {7231} (\bibinfo {year} {1991})}\BibitemShut {NoStop}%
\bibitem [{\citenamefont {Zhao}\ \emph {et~al.}(2008)\citenamefont {Zhao},
  \citenamefont {Morniroli}, \citenamefont {Legris}, \citenamefont {Ambard},
  \citenamefont {Khin}, \citenamefont {Legras},\ and\ \citenamefont
  {Blat-Yrieix}}]{zhao2008identification}%
  \BibitemOpen
  \bibfield  {author} {\bibinfo {author} {\bibfnamefont {Z.}~\bibnamefont
  {Zhao}}, \bibinfo {author} {\bibfnamefont {J.-P.}\ \bibnamefont {Morniroli}},
  \bibinfo {author} {\bibfnamefont {A.}~\bibnamefont {Legris}}, \bibinfo
  {author} {\bibfnamefont {A.}~\bibnamefont {Ambard}}, \bibinfo {author}
  {\bibfnamefont {Y.}~\bibnamefont {Khin}}, \bibinfo {author} {\bibfnamefont
  {L.}~\bibnamefont {Legras}}, \ and\ \bibinfo {author} {\bibfnamefont
  {M.}~\bibnamefont {Blat-Yrieix}},\ }\href {\doibase
  10.1111/j.1365-2818.2008.02136.x} {\bibfield  {journal} {\bibinfo  {journal}
  {J. Microsc.}\ }\textbf {\bibinfo {volume} {232}},\ \bibinfo {pages} {410}
  (\bibinfo {year} {2008})}\BibitemShut {NoStop}%
\bibitem [{\citenamefont {Domain}\ \emph {et~al.}(2002)\citenamefont {Domain},
  \citenamefont {Besson},\ and\ \citenamefont {Legris}}]{domain2002atomic}%
  \BibitemOpen
  \bibfield  {author} {\bibinfo {author} {\bibfnamefont {C.}~\bibnamefont
  {Domain}}, \bibinfo {author} {\bibfnamefont {R.}~\bibnamefont {Besson}}, \
  and\ \bibinfo {author} {\bibfnamefont {A.}~\bibnamefont {Legris}},\ }\href
  {\doibase 10.1016/S1359-6454(02)00173-8} {\bibfield  {journal} {\bibinfo
  {journal} {Acta Mater.}\ }\textbf {\bibinfo {volume} {50}},\ \bibinfo {pages}
  {3513} (\bibinfo {year} {2002})}\BibitemShut {NoStop}%
\end{thebibliography}%

\end{document}